\documentclass[prd,superscriptaddress,twocolumn,nopreprintnumbers,floatfix,nofootinbib, reprint]{revtex4}

\usepackage{textcomp}
\usepackage{amssymb}
\usepackage{amsmath}
\usepackage{graphicx}
\usepackage{dcolumn}
\usepackage{color,units}
\usepackage{lineno}
\usepackage{xspace}
\usepackage{mathtools}
\usepackage{physics}
\usepackage{tensor}
\usepackage{subcaption}
\usepackage{aas_macros}
\usepackage[dvipsnames]{xcolor}
\usepackage[normalem]{ulem}
\usepackage[colorlinks=true,linkcolor=blue,citecolor=blue,urlcolor=blue]{hyperref}%

\newcommand{\citeg}[1]{\citep[e.g.,][]{#1}}

\newcommand{\SPA}{School of Physics and Astronomy, Monash University, Vic 3800, Australia}
\newcommand{\OzGravMonash}{OzGrav: The ARC Centre of Excellence for Gravitational Wave Discovery, Clayton VIC 3800, Australia}
\newcommand{\OzGravSwin}{OzGrav: The ARC Centre of Excellence for Gravitational Wave Discovery, Hawthorn VIC 3122, Australia}
\newcommand{\Swin}{Centre for Astrophysics and Supercomputing, Swinburne University of Technology, Hawthorn VIC 3122, Australia}
\newcommand{\Cardiff}{School of Physics and Astronomy, Cardiff University, Cardiff, CF24 3AA, United Kingdom}

\newcommand{\fs}{f_{\rm{s, BNS}}}
\newcommand{\fd}{f_{\rm{s, NSBH}}}
\newcommand{\epsb}{\eta_{\rm{ BNS}}}
\newcommand{\epsn}{\eta_{\rm{ NSBH}}}
\newcommand{\mtov}{M_{\rm{TOV}}}
\newcommand{\thj}{\theta_{\rm j}}
\newcommand{\thc}{\theta_{\rm core}}

\newcommand{\thjns}{\theta_{\rm j, BNS}}

\newcommand{\thjbh}{\theta_{\rm j, NSBH}}

\newcommand{\risco}{r_{\rm{ISCO}}}
\newcommand{\rdis}{r_{\rm{disrupt}}}
\newcommand{\rbns}{\mathcal{R}_{\rm{BNS}}}
\newcommand{\rsgrb}{\mathcal{R}_{\rm{SGRB}}}
\newcommand{\rnsbh}{\mathcal{R}_{\rm{NSBH}}}
\newcommand{\un}{\rm{Uniform}}
\newcommand{\currentthj}{\thj{}={15^\circ}^{+4^\circ}_{-5^\circ}}
\newcommand{\currentthjnsbh}{\thjbh{}={15.3^\circ}^{+4.0^\circ}_{-5.1^\circ}}
\newcommand{\currentthjbns}{\thjns{}={14.8^\circ}^{+4.0^\circ}_{-5.1^\circ}}
\newcommand{\currentrbns}{\rbns{}=\unit[384^{+431}_{-213}]{\rm{Gpc}^{-3} \rm{yr}^{-1}}}
\newcommand{\currentrnsbh}{\rnsbh{}=\unit[132^{+109}_{-70}]{\rm{Gpc}^{-3} \rm{yr}^{-1}}}

\newcommand{\gpcyr}{\rm{Gpc}^{-3} \rm{yr}^{-1}}

\begin{document}
\title{
Linking the rates of neutron star binaries and short gamma-ray bursts.
}
\author{Nikhil Sarin}
\email{nikhil.sarin@su.se}
\affiliation{\SPA}
\affiliation{\OzGravMonash}
\affiliation{Nordita, KTH Royal Institute of Technology and Stockholm University Roslagstullsbacken 23, SE-106 91 Stockholm, Sweden}
\affiliation{The Oskar Klein Centre, Department of Physics, Stockholm University, AlbaNova, SE-106 91 Stockholm, Sweden}
\author{Paul D. Lasky}
\affiliation{\SPA}
\affiliation{\OzGravMonash}
\author{Francisco Hernandez Vivanco}
\affiliation{\SPA}
\affiliation{\OzGravMonash}
\author{Simon P. Stevenson}
\affiliation{\Swin}
\affiliation{\OzGravSwin}
\author{Debatri Chattopadhyay}
\affiliation{\Swin}
\affiliation{\OzGravSwin}
\affiliation{\Cardiff}
\author{Rory Smith}
\author{Eric Thrane}
\affiliation{\SPA}
\affiliation{\OzGravMonash}

\date{\today}

\begin{abstract}
Short gamma-ray bursts are believed to be produced by both binary neutron star (BNS) and neutron star-black hole (NSBH) mergers. 
We use current estimates for the BNS and NSBH merger rates to calculate the fraction of observable short gamma-ray bursts produced through each channel.
This allows us to constrain merger rates of BNS to $\currentrbns{}$ ($90\%$ credible interval), a $16\%$ decrease in the rate uncertainties from the second LIGO--Virgo Gravitational-Wave Transient Catalog, GWTC-2. 
Assuming a top-hat emission profile with a large Lorentz factor, we constrain the average opening angle of gamma-ray burst jets produced in BNS mergers to $\approx 15^\circ$. 
We also measure the fraction of BNS and NSBH mergers that produce an observable short gamma-ray burst to be $0.02^{+0.02}_{-0.01}$ and $0.01 \pm 0.01$, respectively and find that $\gtrsim 40\%$ of BNS mergers launch jets (90\% confidence).
We forecast constraints for future gravitational-wave detections given different modelling assumptions, including the possibility that BNS and NSBH jets are different. 
With $24$ BNS and $55$ NSBH observations, expected within six months of the LIGO-Virgo-KAGRA network operating at design sensitivity, it will be possible to constrain the fraction of BNS and NSBH mergers that launch jets with $10\%$ precision.
Within a year of observations, we can determine whether the jets launched in NSBH mergers have a different structure than those launched in BNS mergers and rule out whether $\gtrsim 80\%$ of binary neutron star mergers launch jets. 
We discuss the implications of future constraints on understanding the physics of short gamma-ray bursts and binary evolution. 
\end{abstract}.

\maketitle
\section{Introduction}
\label{sec:intro}
Short gamma-ray bursts have long been thought to be associated with the merger of two compact objects \citep{Eichler:1989Natur,Narayan:1992ApJL,Mochkovitch:1993Natur}. 
This was spectacularly confirmed with the coincident gravitational-wave and electromagnetic observations of binary neutron star (BNS) merger GW170817/GRB170817A~\citep{LIGOScientific:2017vwq,LIGOScientific:2017zic}. 
However, BNS mergers are probably not the only progenitors of short gamma-ray bursts with neutron star-black hole (NSBH) mergers likely contributing to the total rate~\citeg{Mochkovitch:1993Natur,Janka:1999ApJL, Barbieri2019}.
Observations of the host galaxy properties and galaxy offsets of short gamma-ray bursts already provide tantalising hints towards this dual population~\cite{Troja2008, Siellez2016, Gompertz2020}. 
However, the relative fraction produced by each progenitor type is unknown.

The total rate density of short gamma-ray bursts observed in the Universe can be written as
\begin{equation}\label{eq:one_to_rule_them_all}
\rsgrb{} = \fs{}\epsb \rbns{} + \fd{}\epsn \rnsbh{}.
\end{equation}
Here $\rbns{}$ and $\rnsbh{}$ are the BNS and NSBH merger rate densities, respectively, $\fs{}$ is the fraction of binary neutron star mergers that successfully launch a jet and $\epsb$ is the ``beaming fraction'' of the jets launched in BNS mergers that are detectable at Earth.
Similarly, $\fd{}$ is the fraction of NSBH mergers that disrupt sufficient matter to produce a short gamma-ray burst jet, and $\epsn$ is the beaming fraction of the jets launched in NSBH mergers. 
Each of these terms is dependent on properties of the population of binaries such as the distribution of progenitor masses and spins, the nuclear equation of state, and physics dictating jet formation, propagation and structure. We discuss the physics of each term in detail in Sec.~\ref{sec:physics}. 
We note that in our definition, $\rsgrb{}$ does \textit{not} include, for example, magnetar flares that are often mistaken for short gamma-ray bursts~\citeg{burns21}. In principle, this can be taken into account by adding another term to Eq.~\ref{eq:one_to_rule_them_all}, however excluding it implicitly assumes we can distinguish magnetar flares from gamma-ray bursts produced by compact binary mergers.

Equation~\ref{eq:one_to_rule_them_all} relates three rate densities in the local Universe. In reality, all three evolve with redshift. Measurements of the BNS and NSBH rate densities are made through gravitational-wave observations in the local Universe with $z\ll1$~\citeg{Abbott2018}, implying the redshift dependence can be safely ignored or assumed to be negligible. 
In contrast, the rate density of short gamma-ray bursts is constrained using observations of short gamma-ray bursts with redshift measurements~\citep{Coward2012} or modelling of the short gamma-ray burst luminosity function~\citep{Wanderman2015, Zhang2018}. Either method introduces a systematic uncertainty up to a factor of $2-3$ on constraints on the short gamma-ray burst rate density in the local Universe~\citep{Coward2012, Wanderman2015, Guo}. We ignore this systematic uncertainty in this Paper, but discuss its implications in Sec.~\ref{sec:conclusion}.

The LIGO-Virgo collaborations recently presented~\citep{LIGOScientific:2021qlt} the first confident observations of gravitational waves from two NSBH coalescences: GW200105\_162426 and GW200115\_042309---throughout this work we refer to these as GW200105 and GW200115, respectively.
Based on the properties of both binaries, \citet{LIGOScientific:2021qlt} calculated the expected dynamical ejecta and mass lost in disk winds, marginalising over the uncertainty in the neutron star equation of state, to determine the disruption probability~\citeg{Foucart_2018,Kruger_2020} of both events. They found that no mass was ejected for either binary at $>99.9$\% probability. That is, neither of these two binaries should have produced a short gamma-ray burst or kilonova~\citep{LIGOScientific:2021qlt}, which is consistent with the lack of observed electromagnetic counterpart~\citep{zhu21_em, Dichiara2021, Anand2021}. 

GW200105 and GW200115 provide us with the opportunity to update our understanding of the astrophysical properties of NSBH systems, including constraints on the merger rate, formation channels of NSBHs, and cosmology~\citeg{Schutz:1986Natur,Vitale:2018PhRvL,Feeney:2021PhRvL,Broekgaarden:2021hlu}. 
If we assume that GW200105 and GW200115 are representative of the NSBH population, their merger rate is constrained to \unit[$45^{+75}_{-33}$]{Gpc$^{-3}$yr$^{-1}$}, and if we account for a broader NSBH population the merger rate is constrained to \unit[$130^{+112}_{-69}$]{Gpc$^{-3}$yr$^{-1}$}~\citep{LIGOScientific:2021qlt}. 
These merger rates suggest that GW200105 and GW200115 are likely to have formed in isolated binaries or young clusters~\citeg{Belczynski_2002,Belczynski_2006,Broekgaarden_2021,Broekgaarden:2021hlu, Mandel2021_review}. Independently, observations of BNS coalescences with aLIGO and Virgo have constrained the rate of BNS mergers in the local Universe to \unit[$320^{+490}_{-240}$]{Gpc$^{-3}$yr$^{-1}$}~\citep{LIGOScientific:2020kqk}\footnote{As this paper was being prepared, LIGO--Virgo published updated BNS and NSBH rate density estimates: $\unit[13-1900]{Gpc^{-3}yr^{-1}}$ and $\unit[7.4-320]{Gpc^{-3}yr^{-1}}$ (90\% credibility), respectively~\cite{gwtc-3_pop}. The increase in uncertainty comes from a more thorough investigation of systematics related to the unknown BNS mass distribution.}.
Observations with gamma-ray observatories have constrained the observed rate of short gamma-ray bursts (i.e., only those beamed towards Earth) in the local Universe to \unit[$8^{+5}_{-3}$]{Gpc$^{-3}$yr$^{-1}$}~\citeg{Coward2012, Wanderman2015}. 

In this Paper, we use these inferred local rates to investigate what constraints can be placed on the fraction of BNS and NSBH mergers that produce short gamma-ray bursts. We explore the physics that dictates these fractions, and forecast constraints that can be obtained in the next few years from gravitational-wave observations as the network sensitivity of detectors improves. 
We explore the implications of such constraints on short gamma-ray burst physics and neutron star binary astrophysics. 

The Paper is laid out as follows. In Sec.~\ref{sec:physics} we expand on Eq.~\ref{eq:one_to_rule_them_all} explicitly detailing the different dependencies and physics of each term. We also calculate each term using different assumptions about the launching and propagation of gamma-ray burst jets, population synthesis outputs, and the nuclear equation of state. 
In Sec.~\ref{sec:currentconstraints} we evaluate Eq.~\ref{eq:one_to_rule_them_all} with minimal assumptions to calculate constraints on the fraction of BNS and NSBH mergers that produce short gamma-ray bursts given current rate estimates. 
In Sec.~\ref{sec:distribution} we explore the assumption that there is no universal jet opening angle but rather a distribution and test whether this can bias our results.
In Sec.~\ref{sec:forecasts}, we discuss the implications future constraints on each term will have on fundamental questions in binary evolution and gamma-ray burst astrophysics. 
We summarise our results and provide concluding thoughts in Sec.~\ref{sec:conclusion}.

\section{Physics of each term}\label{sec:physics}
Equation~\ref{eq:one_to_rule_them_all} connects the BNS and NSBH merger rate to the observed rate of short gamma-ray bursts. Each term is rich in phenomenology and is connected to binary and gamma-ray burst physics. 
In this Section, we provide details on each of these terms, their explicit dependence on fundamental parameters, and calculate these terms based on different assumptions.

\subsection{Jet-launching fractions}
Numerical simulations suggest only BNS mergers whose remnants promptly collapse into a black hole or produce a hypermassive neutron star~\cite[see][for a review on BNS post-merger remnants]{Sarin2021}, can launch jets that power short gamma-ray bursts~\citeg{margalit15_supramassive,murguia17, ciolfi18_sgrb}. 
This constraint implies that any binary resulting in a remnant of mass $\lesssim1.2 \mtov{}$ does not produce a short gamma-ray burst. Here, $\mtov{}$ is the maximum, non-rotating neutron star mass known as the Tolman-Oppenheimer-Volkoff mass; a property of the unknown nuclear equation of state. Explicitly, this implies that $\fs{}$ can be expressed as
\begin{align}
    \fs = \int dm_1 \int dm_2 \,
    p_\textrm{BNS}(m_1, m_2 | M_\text{TOV})
    \pi(m_1, m_2).
\end{align}
Here $p_\textrm{BNS}$ is the probability of a successful jet launch, $\pi(m_1, m_2)$ is the astrophysical distribution of primary and secondary mass.

This relatively straightforward functional dependence has two caveats. 
First, X-ray plateaus of short gamma-ray bursts suggest that many gamma-ray bursts are powered by neutron star central engines~\citeg{rowlinson13, sarin20}, implying the $1.2 \times \mtov{}$ constraint may be incorrect. 
Second, whether a jet is successful in producing prompt gamma-ray emission depends on the Lorentz factor of the jet and its threshold for producing prompt gamma-ray emission, both of which are unknown~\citep{rhoads03, lamb_kobayashi16}. Therefore, even when jets are successfully launched, they may not all produce prompt gamma-ray emission. 
We return to these two caveats in Sec.~\ref{sec:forecasts} and \ref{sec:conclusion}. 

The question of whether an NSBH merger produces a short gamma-ray burst depends on whether enough matter is disrupted to create an accretion disk of sufficient mass around the remnant black hole. 
The mass in the disk must be large enough to feed a relativistic jet, and power it long enough so that it can break through the merger ejecta~\citeg{Foucart_2020}. 
In general, there are two outcomes of an NSBH merger that depend primarily on the progenitor binary's mass ratio and the black hole spin: 1) the neutron star plunges into the black hole without being tidally disrupted, or 2) the neutron star is disrupted. 
To good approximation, disruption occurs when the disruption radius $\rdis{}$ is greater than the black hole's innermost stable circular orbit (ISCO) $\risco{}$, i.e. $\rdis{} \geq \risco{}$. 

The disruption radius can be approximated by (e.g. \cite{Lee_1999,Wiggins_2000})
\begin{equation}\label{eq:r_dis}
    r_\text{disrupt} = k R_\text{NS} \left( \frac{M_\text{BH}}{M_\text{NS}} \right)^{1/3}.
\end{equation}
We assume $k=1$, consistent with results found in the literature~\cite{Lee_1999,Wiggins_2000}.

If the black hole spin is aligned with the orbital angular momentum, the ISCO radius can be approximated in terms of the black hole's mass $M_{\rm BH}$ and dimensionless spin $\chi_{\rm BH}$ as
\begin{equation}\label{eq:r_isco}
    r_\text{ISCO} = f(\chi_\text{BH}) \frac{G M_\text{BH} }{c^2}.
\end{equation}
Here, $ 6 \leq f(\chi_\text{BH}) \leq 9 $ for a co-rotating orbit, and $ 1 \leq f(\chi_\text{BH}) \leq 6 $ for a counter-rotating orbit~\cite{Bardeen_1972}. From Eqs. (\ref{eq:r_dis}) and (\ref{eq:r_isco}), we see that disruption is favoured for low mass ratios $q= M_\text{NS}/ M_\text{BH}$ and high, co-rotating black hole spins. 
The disruption fraction $\fd{}$ can be calculated if we know the population distribution of NSBH progenitor masses and spins, as well as the nuclear equation of state that relates the neutron star progenitor mass to its radius. 
In other words, this disruption fraction can be written functionally as 
\begin{multline}
\fd{} = \int dm_{\text{NS}} \int dm_{\text{BH}} \int d\chi_\text{BH} 
    \ p_\text{NSBH}
    \\ \pi(m_{\text{NS}}, m_{\text{BH}}, \chi_{\text{BH}}),
\end{multline}
where 
\begin{align}
    p_\text{NSBH} = \Theta\left(
    R_\text{NS} \left( \frac{M_\text{BH}}{M_\text{NS}} \right)^{1/3}
    -
    f(\chi_\text{BH}) \frac{G M_\text{BH} }{c^2}
    \right),
\end{align}
is the probability of a successful jet launch in an NSBH  merger (i.e., a Heaviside step function that evaluates to $0$ when $r_\text{disrupt}>r_\text{ISCO}$), and $\pi(m_{\text{NS}}, m_{\text{BH}}, \chi_{\text{BH}})$ is the astrophysical distribution of neutron star and black hole masses and black hole spins of the components that participate in NSBH mergers. 

\begin{figure}
    \centering
    \includegraphics[width=0.48\textwidth]{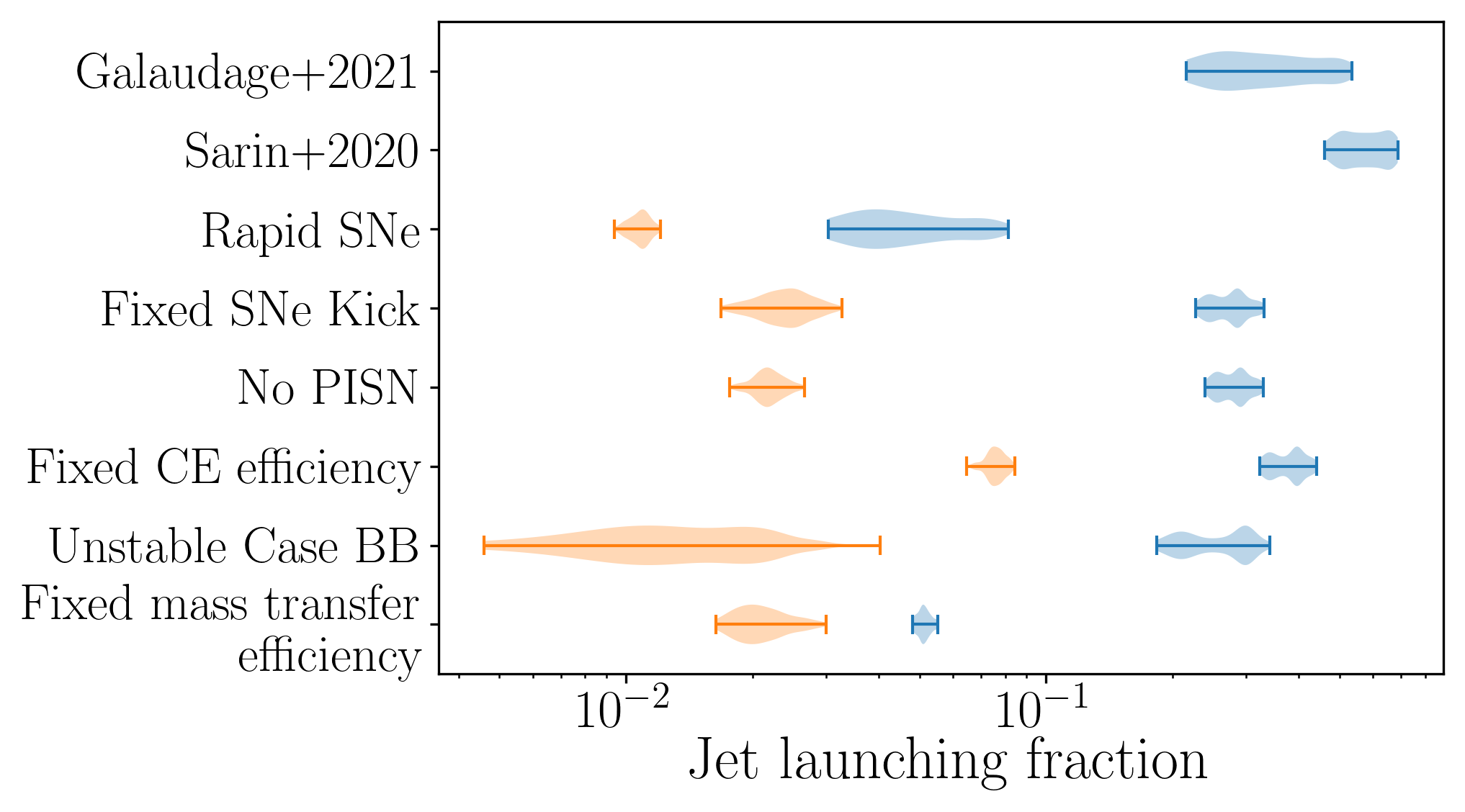}
    \caption{Violin plot of the jet-launching fractions $\fs{}$ (blue) and $\fd{}$ (orange) for different {\sc{compas}} simulations from~\citet{Broekgaarden_2021} built with varying binary evolution physics. For BNS mergers, we also show $\fs{}$ for mass distributions from~\citet{sarin20} with a mixing fraction of GW170817-like and GW190425-like mergers of $50\%$, and the mass distribution inferred from the population of double neutron stars seen in radio and gravitational waves by~\citet{Galaudage2021}. The posterior widths represents the uncertainty due to the unknown nuclear equation of state. Note that we plot the NSBH fraction without the constraint for a minimum disk mass, which would reduce these fractions, although the change would be smaller than that due to the equation of state.}
    \label{fig:jet-launching-fractions} 
\end{figure}

We calculate $\fs{}$ and $\fd{}$ using the above prescription and a suite of results from~\citet{Broekgaarden_2021} that implements different population-synthesis prescriptions for the binary evolution physics---see that Paper, in particular their Tables~1 and 2, for a description of each of the models. 
From these various models, we take the mass distribution of neutron stars and black holes that participate in NSBH and BNS mergers and the orbital period before the second binary component explodes as a supernova, the latter informing the spin distribution of black holes that participate in NSBH mergers~\citep{Qin2018,Broekgaarden_2021}. 
We also marginalize over the unknown equation of state. 
In particular, we use piecewise polytropic equations of state, with polytropic indices uniformly distributed between 1.1 and 4.5, and neutron star central pressure between $10^{32.6}$ and $\unit[10^{33.6}]{dyne/cm^2}$ (see Ref.~\cite{HernandezVivanco2020} for an explanation and implementation of these parameter choices), and the additional constraint that $M_{\rm TOV}$ is between 2.2 and $\unit[2.4]{M_\odot}$ consistent with combined constraints from multimessenger observations of GW170817, GW190425, and NICER observations~\citep{Raaijmakers2021}. We also assume $0.05 M_\odot$ ejecta in every BNS merger, consistent with observations of AT2017gfo~\citeg{Smartt2017}. 
For $\fd{}$ we also explore the affect of an additional constraint; that the disk mass from an NSBH merger must be above $\unit[0.075]{M_\odot}$ to ensure the jet can successfully break through the ejecta as hypothesised in e.g., \citet{Zappa2019}. We calculate $\fd{}$ enforcing this constraint using relations for the ejecta and disk mass in~\citet{Bhattacharya2019}. 
We find that the discrepancy from the additional constraint of a minimum disk mass creates a change in $\fd{}$ smaller than due to the unknown equation of state and we therefore ignore this additional constraint in subsequent sections. 
The results where only disruption is necessary for $\fd{}$ are shown in Fig.~\ref{fig:jet-launching-fractions}. 

Figure~\ref{fig:jet-launching-fractions} shows the jet-launching fractions $\fd{}$ (orange) of NSBH mergers respectively for a diverse set of {\sc{compas}} simulations~\citep[][]{Broekgaarden_2021} with various modifications. In particular, the rapid supernovae model (Rapid SNe), fixed supernovae kick (fixed SNe kick), no pulsational pair instability supernovae (No PISN), fixed common envelope efficiency (Fixed CE efficiency), unstable case BB mass transfer, and fixed mass transfer efficiency. For a full description of these simulations we refer the Reader to \citep{Broekgaarden_2021}.
In general, the different models predict that anywhere between $\approx 0.5-10 \%$ of NSBH mergers disrupt sufficient matter to launch a jet accounting for the unknown equation of state. The smallest disruption values are found for the rapid supernova model~\citep{Fryer2012}, which enforces a mass gap between neutron stars and black holes~\citep[cf.][]{Ozel:2010ApJ,Farr:2011ApJ}. 
This mass gap combined with the relative lack of highly spinning black holes in our model~\citep{Qin2018,Chattopadhyay_2021,Broekgaarden_2021} makes it much more difficult to tidally disrupt the neutron star leading to lower values of $\fd$.

We also calculate the BNS jet-launching fraction $\fs{}$ (blue) given the constraint that the remnant mass $M_{\rm{rem}} \gtrsim 1.2\mtov{}$ for the same set of {\sc{compas}} simulations and two phenomenological distributions based on observations of neutron stars in our Galaxy using radio and gravitational waves~\citep{alsing18, sarin20, Galaudage2021}. In particular, we use the mass distributions from \citet{sarin20} with a mixing fraction of GW170817-like and GW190425-like mergers of $50\%$, and the mass distribution inferred from the population of double neutron stars seen in radio and gravitational waves by~\citet{Galaudage2021}. 

Most models predict $\fs{}$ between $20-60\%$ apart from the {\sc{compas}} models with fixed mass transfer efficiency and rapid supernovae \citep[][]{Fryer2012}, which predict $\fs{} = {5.1}_{-0.2}^{+0.2} \%$ and $\fs{} = {4.9}_{-1.3}^{+2.1} \%$, respectively.  
That is, in these models only $5\%$ of BNS mergers launch a jet. 
The former model fixes the mass transfer efficiency to $0.5$, which leads to significantly lower quantities of heavy double neutron star systems, while the latter as mentioned above enforces a mass gap between neutron stars and black holes. 
In both models, mergers are more likely to form long-lived neutron star remnants which, within the jet-launching model we use (motivated by numerical results~\citeg{murguia17, ciolfi18_sgrb}), are unable to produce short gamma-ray bursts. 
Such a low success rate is unlikely to be correct~\citeg{margalit19, Beniamini2019}, which may hint at either of these models being incorrect or that the assumption of requiring a black hole to launch a jet capable of producing a short gamma-ray burst is flawed. We return to this point in Sec.~\ref{subsec:fs!=1} and Sec.~\ref{sec:conclusion}.
Apart from these two models, all models predict a similar $\fs{}$ between $20-60\%$. However, even a value of $60\%$ is potentially problematic in light of constraints from GW170817, which suggests that most BNS mergers should produce short a gamma-ray bursts~\citeg{Beniamini2019}. 

Binary population-synthesis models typically predict a significant number of BNS with higher masses than those seen in the Galactic population~\citep[][]{Vigna-Gomez:2018dza,Chattopadhyay:2019xye,Mandel:2021ewy}. 
This discrepancy may be due to selection effects that prevent heavier neutron stars (as in GW190425 \citep{LIGOScientific:2020aai}) from being observable in radio~\citeg{Galaudage2021,Vigna-Gomez:2021oqy} or may hint at a systematic issue with population-synthesis results with regards to determining neutron star masses~\citep[][]{Mandel:2021ewy}. 
Our constraint that immediate formation of a black hole is required to launch a jet (i.e., $M_{\rm{rem}} \lesssim1.2 \times \mtov{}$), implies that population-synthesis results predict a higher $\fs{}$ than one predicted by the distribution of BNS just in our Galaxy. 
For example, taking the distributions derived in~\citet{kiziltan13} and~\citet{farrow19}, we find $\fs{} = {0.01}_{-0.01}^{+0.03}$ and $\fs{} = {0.02}_{-0.01}^{+0.03}$, respectively; such low values are unlikely to be correct. 
It is worth noting that these distributions were derived before the observation of GW190425, and are inconsistent with the masses inferred in that BNS.

\subsection{Beaming fractions}
\label{subsec:beaming_fractions}

The two terms $\fs{}$ and $\fd{}$ dictate the fractions of BNS and NSBH mergers that can produce a short gamma-ray burst. 
A significant (but unknown) fraction of these jets are pointed away from Earth and are therefore not observable or are too dim to be detected by current gamma-ray detectors. 
We denote the observable fractions of short gamma-ray bursts from BNS and NSBH mergers as $\epsb$ and $\epsn$, respectively. 

Historically, the beaming fraction $f_{b}$ has been estimated through the relation $f_{b} = 1 - \cos(\thj{})$~\cite{Frail2001}, where $\thj{}$ is the opening angle of the gamma-ray burst jet\footnote{This is sometimes defined as $f_b=(1 - \cos{\thj{}})^{-1}$, however we use the original definition as presented in Ref.~\citep{Frail2001}}. 
This implicitly assumes that the jet structure is top-hat with a large Lorentz factor such that all gamma-ray bursts observed within the opening angle $\thj{}$ are observable and those outside are not. Constraints on the beaming fraction range from $\sim 10^{-3}-0.015$~\citeg{berger14}, which implicitly assumes that short gamma-ray bursts produced by BNS and NSBH mergers have the same opening angles so that $\epsb=\epsn$. 

In general, we do not expect the beaming fractions of BNS and NSBH jets to be the same given the different jet-launching environment. The beaming fraction is a function of the initial jet structure, jet-launching mechanism, and the propagation of the jet through the surrounding ejecta. 
Typically, BNS mergers have more ejecta than NSBH mergers along the poles~\citeg{Foucart_2020, Kyutoku2021}, which is the region where jets are launched. 
Numerical simulations already show that propagation through ejecta helps collimate gamma-ray burst jets~\cite{Barbieri2019, Urrutia2021}. 
We therefore believe that gamma-ray burst jets launched in NSBH mergers are likely to be less collimated than ones launched in BNS mergers. 
Similarly, depending on binary properties, some NSBH mergers are unable to form a disk of sufficient mass to power a jet long enough to make its way out of the merger ejecta~\citeg{Zappa2019}, which will likely lead to a cocoon-like outflow. 
It is worth noting that the lower amount of overall mass in the polar region may alleviate this concern~\citep{Just2016, Kyutoku2021}.

The above arguments implicitly assume gamma-ray emission is not produced outside the ultra-relativistic core, something we now know to be incorrect following observations of GRB170817A~\cite{matsumoto19}. 
These observations also suggest that other gamma-ray bursts may be viewed from outside the opening angle of the jet, a suggestion that has led to reclassification of some faint short gamma-ray bursts such as GRB150101B as ones viewed off axis~\cite{troja18_150101B}. 
The presence of off-axis gamma-ray bursts makes it difficult to parameterise the beaming fraction as simply being related to the opening angle of the jet. 
To further complicate matters, there is no robust model for calculating the flux of the prompt emission that can robustly predict the vast range of gamma-ray burst energies. 
In light of GRB170817A, one recent approach to estimate the flux of the prompt emission is to compute the peak flux at a given observer viewing angle using the same structured jet profile used to describe the afterglow for a given gamma-ray detector band-pass and threshold~\citeg{howell19, Salafia2019}. 
For our purposes, we use this aforementioned prescription, accompanied by a choice of jet structure, to compute the flux for an isotropic distribution of observers and calculate the fraction of short gamma-ray bursts that are observable with the \textit{Fermi} telescope.

To understand the beaming fractions of BNS $\epsb$ and NSBH $\epsn$, we use two different viewing-angle dependent energy distributions for the jet structures: a Gaussian and a power-law structured jet respectively given by
\begin{align}
    E(\theta)=& E_{0} \exp \left(-\frac{\theta^{2}}{2 \thc^{2}}\right),\label{eq:Gaussian}\\
    E(\theta)=& E_{0}\left(1+\frac{\theta^{2}}{b \thc^{2}}\right)^{-b / 2}.\label{eq:PowerLaw}
\end{align}
Here $E_{0}$ is the on-axis isotropic equivalent energy, $\theta$ is the angle from the jet axis, $\thc{}$ is the half opening angle of the ultra-relativistic core of the jet, and $b$ is the power-law slope. Often these models are parameterised with an additional angle $\theta_{\rm{wing}}$, beyond which the energy is zero, implying no gamma rays can be produced beyond this angle. 
This angle is typically taken to be some integer multiple of $\thc{}$~\citeg{Cunningham2020}.
The parameters of the Gaussian and power-law jet structured models are different for BNS and NSBH as we expect BNS jets to be more collimated, as described in Sec.~\ref{sec:physics}. 

\begin{figure}
    \centering
    \includegraphics[width=0.48\textwidth]{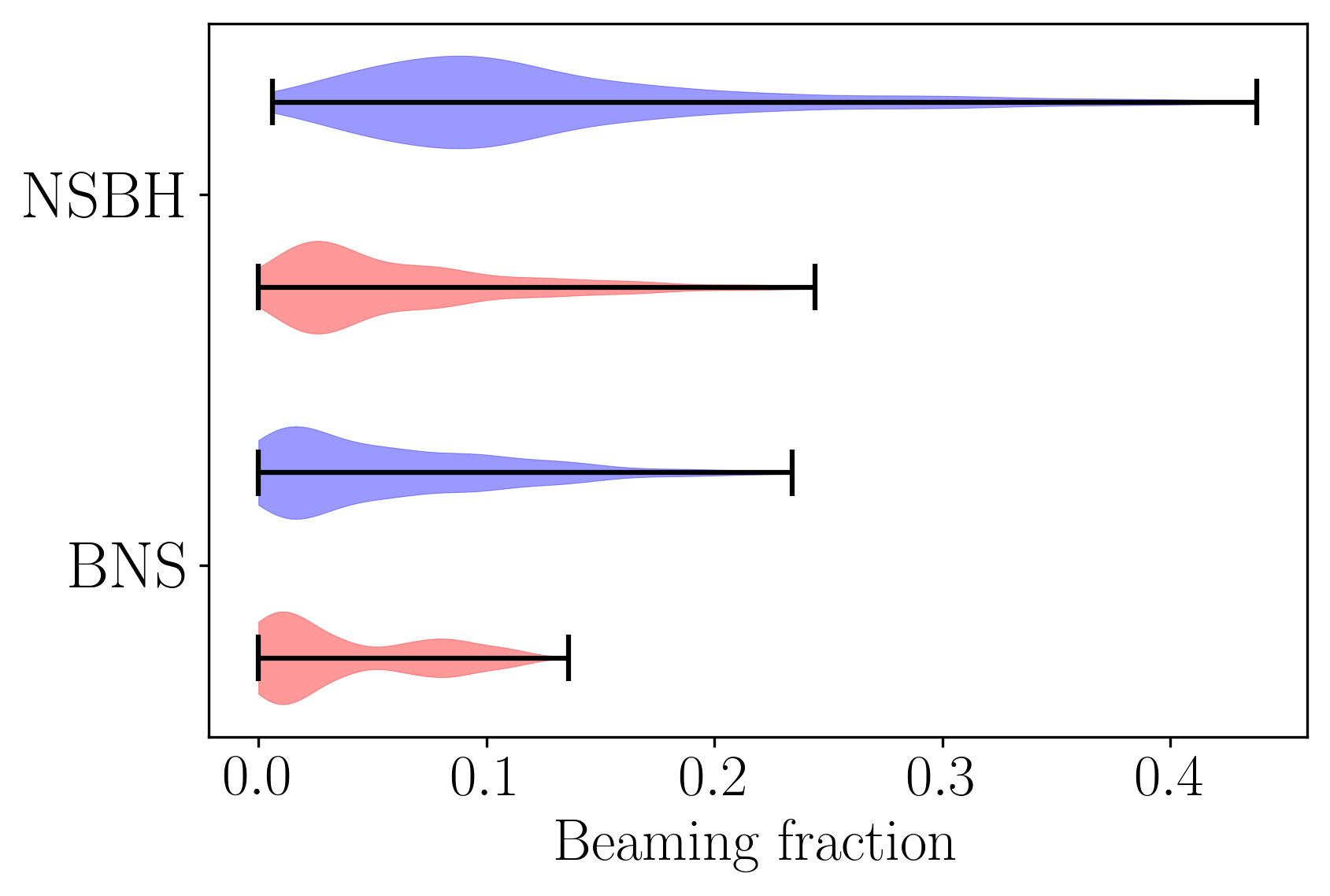}
    \caption{The beaming fractions $\epsb$ and $\epsn$ for a realistic population of NSBH and BNS mergers occurring within their respective horizon distance of aLIGO at design sensitivity, with the viewing angle distributed uniformly in $\cos(\iota)$. The blue violins correspond to the beaming fraction predicted using the power-law jet model while the red violins correspond to the fraction predicted using the Gaussian jet structure. The distribution represents the uncertainty on these fractions due to the unknown gamma-ray burst jet structure parameters.}
    \label{fig:beaming} 
\end{figure}

In Fig.~\ref{fig:beaming} we plot the beaming fractions $\epsb$ and $\epsn$ for a realistic population of BNS and NSBH out to \unit[330]{Mpc} and \unit[590]{Mpc} (corresponding to the horizon distances for aLIGO operating at design sensitivity~\citep{Abbott2018}, where the NSBH horizon distance assumes $\unit[1.4]{M_\odot}$ + $\unit[10]{M_\odot}$ systems), and viewing angles that are uniform in $\cos(\iota)$. The red violins correspond to the Gaussian structured jet while the blue violins correspond to the power-law structured jet. The priors for the BNS jets are motivated by fits to the afterglow of GRB170817A~\citep{Ghirlanda2019} and expectations of structured jets produced in NSBH mergers~\citep{Barbieri2019}. 

Figure~\ref{fig:beaming} shows two main trends: 1) power-law structured jets tend to have higher beaming fractions than gaussian jets and 2) NSBH jets have higher beaming fractions than BNS jets. 
The former is due to the energy distribution of power-law jets as opposed to Gaussian jets, with power-law jets having more energy at higher angles away from the jet axis than Gaussian. 
The latter is a realisation of our prior assumption that NSBH jets are less collimated. 
It is worth noting that the range of beaming fractions predicted by either a Gaussian or power-law structured jet is broad for both BNS and NSBH jets and is an encapsulation of both our uncertainty in the jet-structure parameters and their priors and the unknown prompt emission generation mechanism. This uncertainty will improve with future multi-messenger events like GW170817~\citep{Beniamini2019, Biscoveanu2020, Farah2020}. In general, we find that $\epsb \lesssim 0.1$ and $\epsn \lesssim 0.3$ for the less collimated power-law jet model at $90\%$.

In the following sections, we use these models to estimate constraints on each of these terms given current short gamma-ray burst, BNS and NSBH merger-rate measurements. We also forecast each of these rates assuming future gravitational-wave and electromagnetic observations.
\section{Results}\label{sec:results}
\subsection{Top hat jets}\label{sec:currentconstraints}
As described in Sec.~\ref{sec:physics}, each term in Eq.~\ref{eq:one_to_rule_them_all} is a function of several physical parameters. 
These parameters can be informed through gamma-ray burst modelling and population synthesis studies, as well as inferences on the observed population. 
As a first attempt, we consider constraints imposed by a model with minimal assumptions and directly comparable to historical measurements. 
Our minimal working model assumes the beaming fraction of both BNS and NSBH is the same ($\epsb=\epsn$), and is parameterised only by the jet opening angle i.e., that all gamma-ray bursts are observable as long as they are observed within some opening angle $\thj$. To wit
\begin{multline}
\label{eq:in_the_darkness_find_them}
\rsgrb{} = \left(1-\cos\thj\right)\left(\fs\rbns+\fd\rnsbh\right).
\end{multline}

We place priors directly on $\fs{}$ and $\fd{}$, which uses two pieces of information. 
First, that we expect a large majority of BNS mergers to produce short gamma-ray bursts~\citeg{margalit19, Beniamini2019} and second, that the two bona fide NSBH mergers detected with gravitational waves should not have produced short gamma-ray bursts~\citeg{LIGOScientific:2021qlt}. 
These two statements can be translated into priors as 
\begin{align}
    \pi(\fs)&=\un{}[0.1, 1]\label{eq:prfs},\\
    \pi(\fd)&=(1-p)^{2}.\label{eq:pifd}
\end{align}
Equation~\ref{eq:prfs} is a uniform distribution motivated by observations of GW170817~\citep{Beniamini2019}. Equation~\ref{eq:pifd} is a binomial probability distribution for zero successes from two independent trials, where $p$ is the success probability for each trial. 
We also enforce that $\fs{} \gtrsim \fd{}$ due to theoretical motivations outlined in Sec.~\ref{sec:physics}.

\begin{figure}
    \centering
    \includegraphics[width=0.5\textwidth]{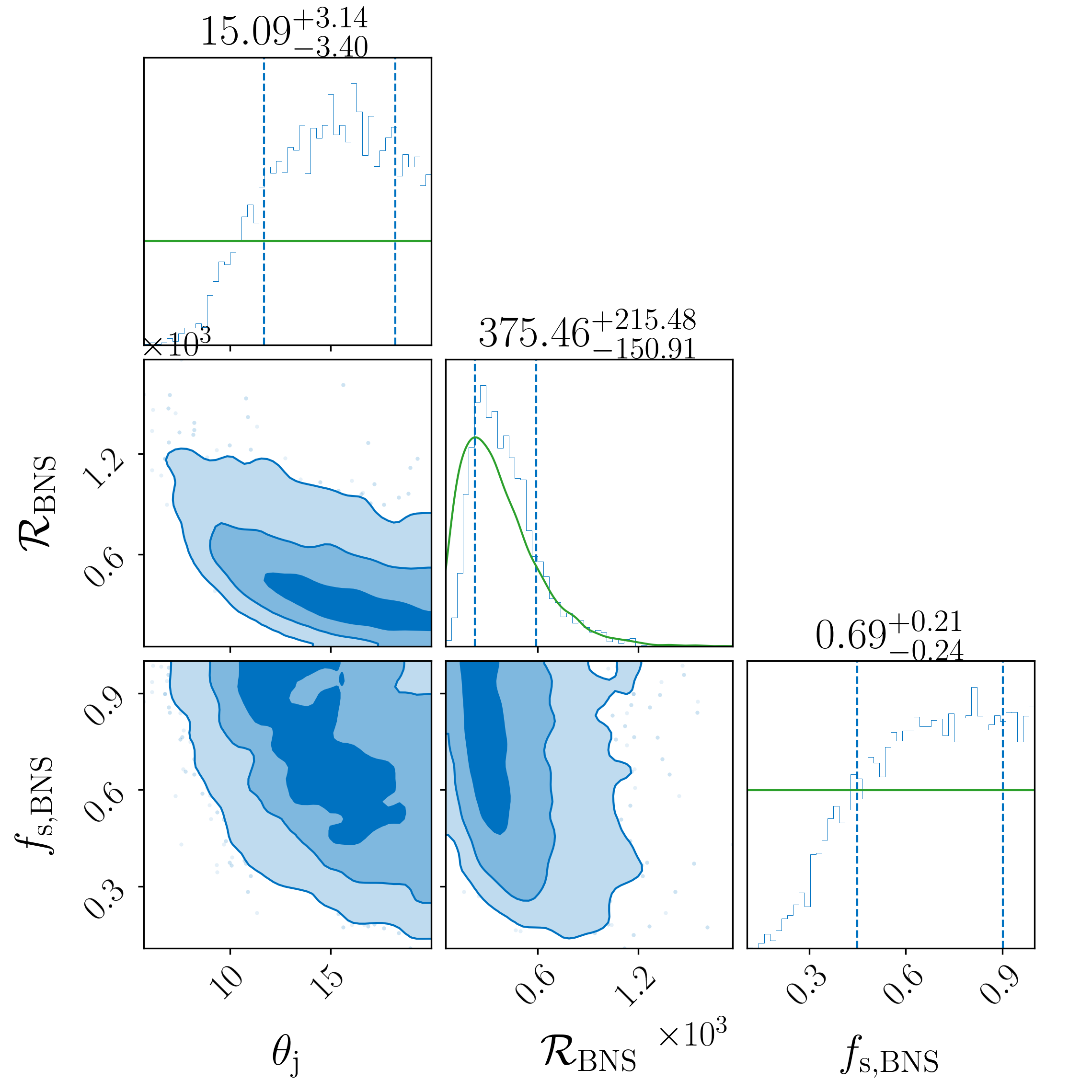}
    \caption{Current constraints on the fiducial model (Eq.~\ref{eq:in_the_darkness_find_them}). The values quoted above are the median and $1\sigma$ credible intervals. The green curves represent the prior while the blue shaded regions represents $1-3\sigma$ credible intervals. The posterior distributions of the other parameters in the model are equivalent to their priors.}
    \label{fig:current_basic_constraints} 
\end{figure}

We evaluate Eq.~\ref{eq:in_the_darkness_find_them} using the above priors and rate measurements $\rbns{} = \unit[320^{+490}_{-240}$]{Gpc$^{-3}$yr$^{-1}$}~\citep{LIGOScientific:2020kqk}, $\rnsbh{} =\unit[130^{+112}_{-69}$]{Gpc$^{-3}$yr$^{-1}$}~\citep{LIGOScientific:2021qlt}, and $\rsgrb{} = \unit[8^{+5}_{-3}$]{Gpc$^{-3}$yr$^{-1}$}~\citeg{Coward2012, Wanderman2015}. 
With the assumption that both NSBH and BNS merger jets have the same opening angle, measurements of the merger and short gamma-ray burst rates place strong constraints on the opening angle of the jet and the BNS merger rate. 
We show this two-dimensional marginalised posterior distribution in Fig.~\ref{fig:current_basic_constraints}.  
We find that the opening angle of gamma-ray burst jets is constrained to $\currentthj{}$ ($90\%$ credible interval), consistent with other values in the literature~\citeg{berger14, Williams2018}. 
Note that we place a hard cut off at $\thj{} = 20^\circ$ as larger opening angles would be inconsistent with constraints on short gamma-ray burst energetics~\citep{Frail2001, Fong2012}.
Similarly, the BNS merger rate is constrained to $\rbns{}=\unit[375^{+436}_{-211}]{\rm{Gpc}^{-3} \rm{yr}^{-1}}$, which is approximately $15\%$ more informative than the second gravitational-wave catalog merger rate estimate~\citep{LIGOScientific:2020kqk}. 
We measure $\fs{}=0.69^{+0.27}_{-0.37}$ ($90\%$ credible interval) and that $\fs \gtrsim 0.3$ with $95\%$ confidence ruling out population synthesis models with rapid supernovae and fixed mass transfer efficiency (see Fig.~\ref{fig:jet-launching-fractions}). 
Within these assumptions, the NSBH merger rate and jet launching fraction is not constrained from the prior.
We also find an anti-correlation between $\thj{}$ and $\rbns{}$ which agrees with physical intuition: small jet opening angles must be compensated by a larger BNS merger rate to explain the measured short gamma-ray burst rate. 

\begin{figure}
    \centering
    \includegraphics[width=0.5\textwidth]{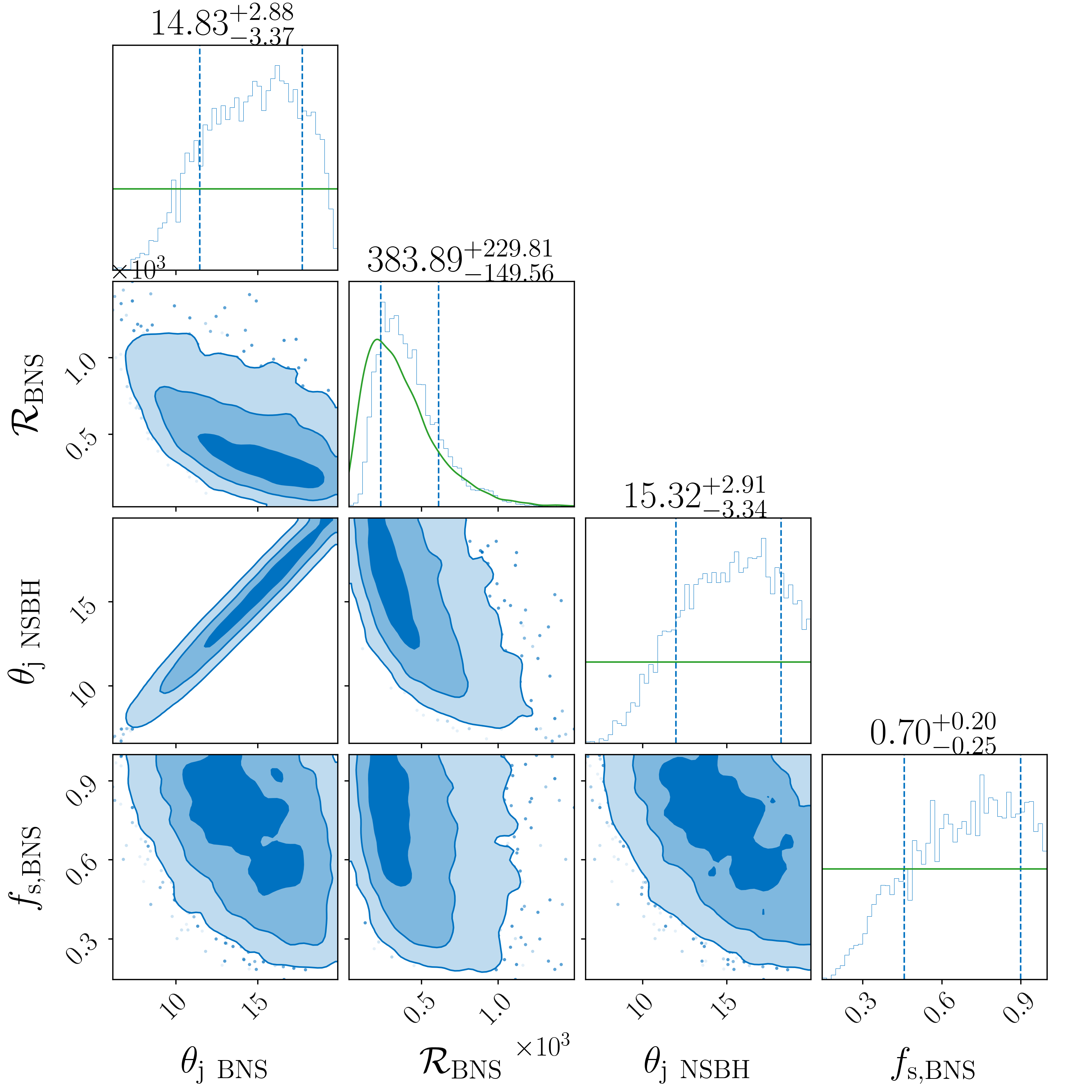}
    \caption{Current constraints on the fiducial model (Eq.~\ref{eq:in_the_darkness_find_them}) but with $\thj{}$, allowed to be different for BNS and NSBH mergers. The values quoted above are the median and $1\sigma$ credible interval. The green curves represent the prior while the blue shaded regions represents $1-3\sigma$ credible intervals.}
    \label{fig:current_two_frac_constraints} 
\end{figure}

We now relax the assumption that jets launched in NSBH mergers have the same opening angle as those in BNS mergers; as described in Sec.~\ref{sec:physics}, we expect the former to be less collimated than the latter. 
We again evaluate Eq.~\ref{eq:in_the_darkness_find_them} using current rate measurements, but now assume $\thj{}$ is different for NSBH and BNS, parameterising both as uniform distributions between $\unit[1-20]{^{\circ}}$ motivated by estimates of the beaming fraction of short gamma-ray bursts and their overall energetics~\citeg{Fong2012, berger14}. The other priors are the same as used above. 
We also enforce $\thjbh{} \gtrsim \thjns{}$ due to theoretical motivations outlined in Sec.~\ref{sec:physics}.
The posterior distribution on a subset of model parameters are shown in Fig.~\ref{fig:current_two_frac_constraints}.

Under this model with fewer assumptions, we measure BNS and NSBH opening angles to $\currentthjbns{}$ and $\currentthjnsbh{}$ respectively. 
These opening angles are effectively the same, and may suggest that within the model where the beaming fractions are given by $1 - \cos{\thj{}}$, there is no significant difference between BNS and NSBH jets. 
This may imply that there is a maximum angle away from the jet axis at which prompt gamma-ray radiation cannot be produced, and that this is the same for both BNS and NSBH jets or that the difference is smaller than we can currently probe.

The analysis presented hitherto also provides updated rate measurements that are slightly improved over current estimates in the literature. 
We measure the BNS merger rate as $\currentrbns{}$ and the NSBH merger rate as $\currentrnsbh{}$ ($90\%$ credible interval), which are approximately $16\%$ and $3\%$ more informative than the BNS and NSBH merger rates from the second gravitational-wave catalog~\citep{LIGOScientific:2020kqk}, respectively.
\subsection{Distribution of opening angles}\label{sec:distribution}
The previous Section used the  $f_{b} = 1 - \cos{\thj{}}$ relation as an estimate of the ``beaming fraction'' to allow direct comparison with historical measurements. 
However, in reality, we do not expect this relation to be correct for two reasons. 1) It is unlikely all gamma-ray bursts have the same jet opening angle and 2) prompt gamma-ray radiation can be produced outside the ultra-relativistic core as has been argued for  GRB170817A~\citep{matsumoto19}. Here we address the impact of ignoring these two issues on our analysis. 

As described in Sec.~\ref{sec:physics}, interaction with the ejecta around the polar region helps collimate jets~\citep{Barbieri2019, Urrutia2021}. 
We do not expect every BNS merger or NSBH merger to eject the same amount of matter so one cannot reasonably expect there to be a universal opening angle $\thj{}$ across multiple BNS mergers, let alone between BNS and NSBH mergers. 
Moreover, both the propagation of a jet and the overall jet structure are strongly connected to the initial jet structure and energetics~\citep{Nativi2021}. 
Both of these are sensitive to binary parameters such as the masses, spins, nuclear equation of state, etc, although it is worth noting that no robust physical relationship exists in the literature linking these variables.

To investigate the effect of assuming a universal jet opening angle on our results above, we perform a more realistic simulation. 
In particular, we rewrite Eq.~\ref{eq:in_the_darkness_find_them} as
\begin{multline}
\label{eq:distribution}
\rsgrb{} = \int d\thj{}\left(1-\cos\thj\right) \\ \times \left(\fs\rbns+\fd\rnsbh\right) \pi(\thj{}),
\end{multline}
where $\pi(\thj{})$ is the astrophysical distribution of jet opening angles relaxing the assumption implicit in Eq.~\ref{eq:in_the_darkness_find_them} of a universal opening angle.
We simulate the short gamma-ray burst rate using Eq.~\ref{eq:distribution} with $\rbns{} = \unit[320]{\gpcyr{}}$, $\rnsbh{}=\unit[130]{\gpcyr{}}$, $\fs{} = 0.3$, $\fd{} = 0.02$ and $\thj{}$ drawn from a normal distribution with mean $\mu = 15^{\circ}$, and $\sigma = 4^{\circ}$. We then perform our analysis by estimating the average opening angle $\theta_{j, avg}$ with uniform priors on $\fs{}$, $\fd{}$, and $\theta_{j, avg}$ between $0.01-0.2$, $0.1-1$, and $1-30^\circ$, respectively. We use the measured merger rates from the second gravitational-wave transient catalog as our priors for $\rbns{}$ and $\rnsbh{}$.  

\begin{figure}
    \centering
    \includegraphics[width=0.5\textwidth]{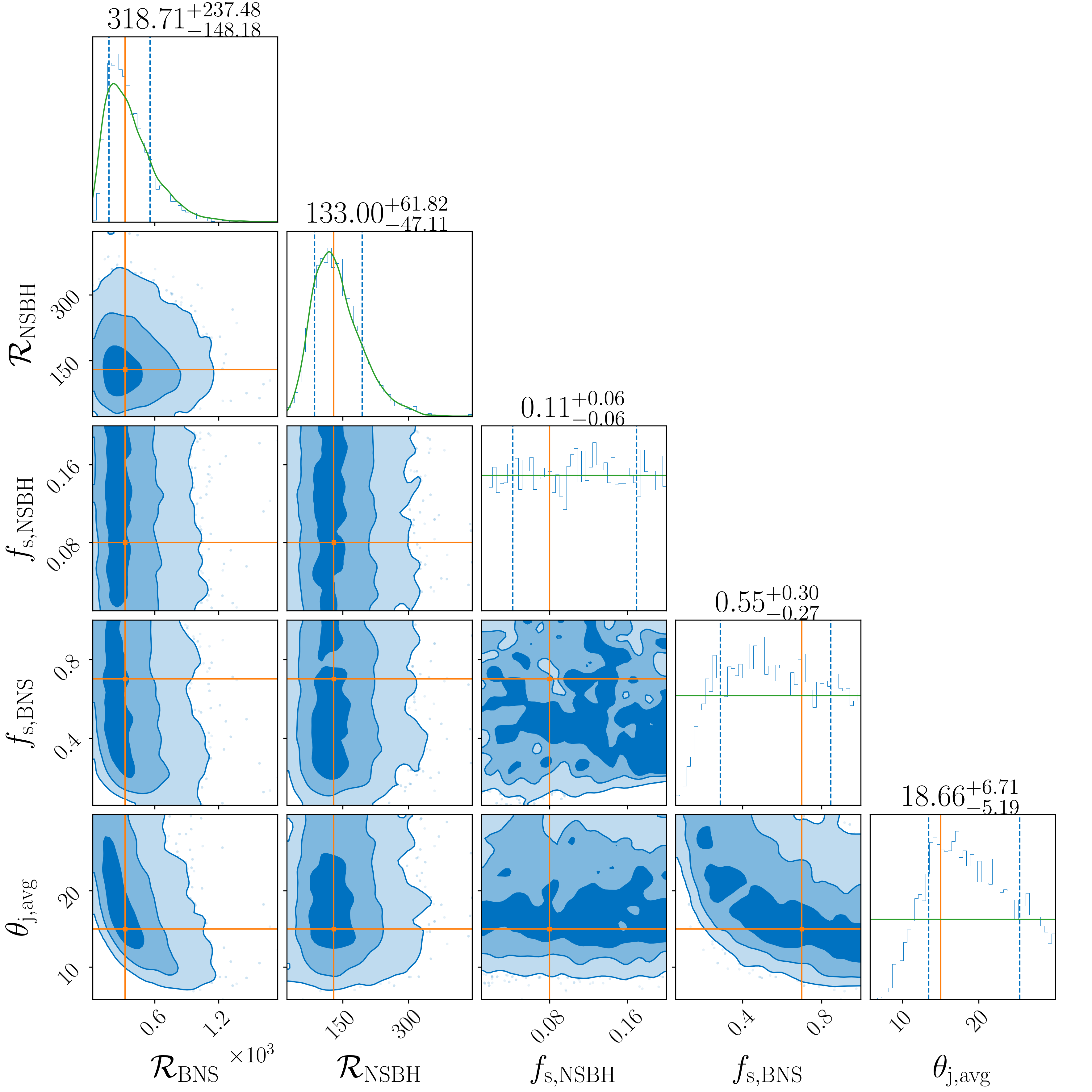}
    \caption{Posterior distribution of all parameters in Eq.~\ref{eq:in_the_darkness_find_them} using simulated data where the jet opening angle is normally distributed rather than taking on a single value. The values quoted above are the median and $1\sigma$ credible intervals. The green curves represent the prior while the blue shaded regions represents $1-3\sigma$ credible intervals. The orange lines represent the values used to simulate the data.}
    \label{fig:opening_angle_dist} 
\end{figure}

In Figure~\ref{fig:opening_angle_dist}, we show the one and two-dimensional posterior distributions of all parameters obtained by fitting the simulated data. 
The orange lines refer to the injected values while the green curves represent the prior. 
All parameters are recovered correctly, with the injected parameter within the $1\sigma$ credible interval, suggesting that given our current range of uncertainty on the rate of NSBH and BNS mergers and short gamma-ray bursts ignoring the distribution of $\thj{}$ does not bias our results. However, we find that the uncertainty on $\theta_{j,avg}$ is larger by $\approx 40\%$ compared to analysis where a single universal jet opening angle is assumed implying that the uncertainties on $\thj{}$ derived in Sec.~\ref{sec:currentconstraints} are underestimated by at least $40\%$. 
\subsection{Structured jets}\label{sec:structuredconstraints}
In the previous Section, we implicitly assumed a ``top-hat" jet structure i.e., that there is a maximum opening angle $\thj{}$ outside of which we observe no gamma-ray emission. As discussed in Sec.~\ref{sec:physics}, we know from observations of GRB170817A, that this is incorrect and that the jet is likely structured.

In this Section, we link the rates of short gamma-ray bursts and neutron star binaries together using Eq.~\ref{eq:one_to_rule_them_all} with priors on $\eta_{\rm BNS}$ and $\eta_{\rm NSBH}$ informed by the Gaussian structured jet model i.e., the beaming fractions shown in Fig.~\ref{fig:beaming} in red. We use the same priors for all other parameters as in Sec.~\ref{sec:currentconstraints} and enforce $\eta_{\rm NSBH} \gtrsim \eta_{\rm NSBH}$ due to the theoretical reasons outlined in Sec.~\ref{sec:physics}.

With current measurements, our constraints with a structured-jet model are similar to the ones obtained with a top-hat model in Sec.~\ref{sec:currentconstraints}. In particular, we measure $\fs{}=0.69^{+0.28}_{-0.37}$ and $\rbns{}=\unit[360^{+407}_{-238}]{\rm{Gpc}^{-3} \rm{yr}^{-1}}$ with $\rnsbh{}$ and $\fd{}$ practically indistinguishable from the prior. This suggests that given current measurements, and the priors used on $\fs{}$ and $\fd{}$, this framework can not probe the difference between a top-hat jet and a Gaussian jet structure. The analysis does provide an improved constraint on the beaming fraction of $\eta_{\rm {BNS}} = 0.03^{+0.05}_{-0.02}$ and $\eta_{\rm {NSBH}} = 0.08^{+0.07}_{-0.05}$ for BNS and NSBH jets, approximately $60\%$ more informative than the prior. However, with structured-jet models the ``beaming fractions" themselves are not particularly enlightening, and one would instead want constraints on the parameters such as the distribution of on-axis isotropic equivalent energy, $E_0$, or the half-opening angle of the ultra-relativistic core, $\thc{}$ that determine the beaming fraction. These constraints require Bayesian hierarchical inference, a development we leave to future work. We note that given current measurements, the constraints on these parameters will not be particularly informative but could become a powerful complementary way to probe the jet structure in the future.
\section{What may the future bring?}\label{sec:forecasts}
In the previous Section, we put priors on $\fs{}$ and $\fd{}$ without incorporating additional binary physics or the nuclear equation of state. We also made the simple historically-motivated assumption that the beaming fraction is solely dependent on $\thj{}$ i.e., we assumed a top-hat jet structure where every gamma-ray burst viewed within some opening angle is observable. 
As described in Sec.~\ref{sec:physics}, each of these terms depends on several other parameters that are functions of binary and gamma-ray burst physics. 
The jet-launching fractions $\fs{}$ and $\fd{}$ depend on the mass distribution of neutron stars that participate in NSBH and BNS mergers, the mass distribution of black holes that participate in NSBH mergers, and their spin distribution. These fractions then also depend on the nuclear equation of state, which dictates the maximum allowed neutron star mass, $\mtov{}$ and the distribution of neutron star radii $R_{\rm NS}$. These dependencies allow us to forecast what we can learn about binary and gamma-ray burst physics as we observe more events.

Current generation gravitational-wave observatories are being upgraded for their fourth observing run \citep{Abbott2018}. These upgrades are expected to bring an increase in sensitivity of a factor 1.5 over the third observing run corresponding to an increase in the observed rate by a factor of approximately $3.5$. Taking the median rate posterior for BNS and NSBH mergers from LIGO/Virgo observations~\cite{LIGOScientific:2020kqk,LIGOScientific:2021qlt}, two years of observation at this assumed sensitivity implies a median expectation value of 18 BNS and 39 NSBH mergers. 
On average, the Neil Gehrels \textit{Swift} Observatory~\citep{swift} observes approximately six to eight short gamma-ray bursts per year. Our median expectation for the number of short gamma-ray bursts over this two-year period is therefore 12 to 16.
The uncertainty on rates decreases roughly as $1/\sqrt{N}$, where $N$ is the number of observations. 
We can use the above relation to forecast how the merger rate, and therefore the terms in Eq.~\ref{eq:one_to_rule_them_all}, will become better constrained in the coming years.

The set of models introduced hitherto are extensive. To answer the most pertinent questions, we reduce this set of models choosing to focus on a limited subset. 
In particular, we use only the Gaussian structured-jet models (Eq.~\ref{eq:Gaussian}) for both BNS and NSBH, and focus only on population-synthesis models for fixed supernovae kick and fixed common-envelope efficiency. 
In the following subsections, we use the jet-launching and beaming fractions motivated by these models to answer the following questions: A) how many observations are required to know whether all BNS mergers launch jets (i.e., rule out $\fs{}=1$), 
B) how many observations are required to say whether the beaming fraction (i.e., the jet structure) of BNS and NSBH are the same (i.e., is $\epsn\neq\epsb$), and C) how many observations are required to determine the difference between different population-synthesis models and therefore learn about binary evolution. In each subsection we perform these calculations, and discuss the physical implications of such results. We note that for all following analysis we simulate data with $\rbns{} = \unit[320]{\gpcyr{}}$, $\rnsbh{}=\unit[130]{\gpcyr{}}$ corresponding to the median merger rate estimates from the second gravitational-wave transient catalog~\citep{LIGOScientific:2020kqk}.

\subsection{Do all BNS launch jets?}
\label{subsec:fs!=1}

Constraining the fraction of BNS that successfully launch a jet ($\fs{}$) has several important implications. As mentioned in Sec.~\ref{sec:physics}, numerical simulations suggest a gamma-ray burst jet is only produced when the remnant mass of a BNS merger is $\gtrsim 1.2 \times \mtov{}$. Depending on the BNS mass distribution, this could imply that a significant fraction of BNS mergers do not produce short gamma-ray bursts, which must be represented in the respective rates. However, X-ray afterglow observations of several short gamma-ray bursts have features that are difficult to explain without requiring a neutron star engine~\citeg{rowlinson13, sarin20}. 
These X-ray observations indicate that $\gtrsim 30\%$ of short gamma-ray bursts produce long-lived neutron stars. Placing the constraint that a jet is only launched if $M_{\rm{rem}} \gtrsim 1.2 \times \mtov{}$,  our different suite of models combined predict a BNS jet-launching fraction $\fs{} = {0.30}_{-0.24}^{+0.31}$ ($90\%$ credible interval). This fraction is lower than expectations in light of GW170817~\citep{margalit19, Beniamini2019}. For example, \citet{Beniamini2019} predicted $\fs{} \approx 0.6-1$ using the short gamma-ray burst luminosity function~\citep{Wanderman2015} and various structured jet models, hinting at a potential inconsistency.

Here, we explore how many observations of BNS and NSBH mergers are necessary to determine whether a black hole central engine is necessary to launch a jet i.e., how long to confidently determine $\fs{} \neq 1$? We create simulated data following Eq.~\ref{eq:one_to_rule_them_all} with $\epsb = 0.02$ and $\epsn = 0.04$ for BNS and NSBH beaming fractions, respectively, and $\fs{} = 0.3$ and $\fd{} = 0.02$ for the jet-launching fractions of BNS and NSBH mergers. 
We estimate the constraints for varying number of observations, simulating new constraints on the merger rates using the current distribution and decreasing the uncertainty with $1/\sqrt{N}$ where $N$ is the number of observations. 
We use broad uninformative priors on the jet-launching fractions and priors informed by the Gaussian jet structure for the beaming fraction. We also enforce the constraint that $\fs{} \gtrsim \fd{}$ and $\epsb \gtrsim \epsn$.

Within a single year of observations at design sensitivity, we expect $48$ and $111$ BNS and NSBH mergers allowing us to confirm if most BNS mergers can successfully launch a jet i.e., rule out $\fs{}\gtrsim 0.8$ with $90\%$ confidence. 
We note that this may not necessarily solve the problem in our understanding of whether short gamma-ray bursts can be produced by mergers producing long-lived neutron stars, as the missing fraction could be gamma-ray burst jets that have low Lorentz factors such that they can not produce prompt gamma-ray emission. However, by this time we may likely have a good understanding of the BNS mass distribution \citep{Chatziioannou:2020msi} such that we could isolate this fraction of `failed' gamma-ray bursts~\citeg{lamb_kobayashi16} from the fraction of remnants with mass $M_{\rm rem} \lesssim 1.2 \times \mtov{}$, providing an answer to one of the most fundamental questions in gamma-ray burst physics.

\subsection{Are all jet structures the same?}
\label{subsec:chi=eta}

The afterglow observations of GRB170817A confirmed that gamma-ray burst jets are structured, but the exact jet structure is unclear. However, more pertinent to our discussion is whether the jet structures of gamma-ray bursts launched in NSBH and BNS mergers are different and ultimately whether they have the same beaming fraction i.e., whether $\epsn = \epsb$. 
As described in Sec.~\ref{sec:physics}, we expect the structures and therefore beaming fractions to be different. This is borne out from numerical simulations which suggest that gamma-ray burst jet structure is a shaped largely by the jet's interaction with the ejecta~\citep{Urrutia2021, Nativi2021} and the expectation that the amount and location of ejecta in BNS and NSBH mergers are different~\citep{Kyutoku2021}.

We use the same procedure and parameters as outlined above to estimate when we will be able to rule out both beaming fractions being equal. 
However, unlike previously, we use priors motivated by the fixed supernovae kick population synthesis model on the jet-launching fraction and broad uninformative priors on $\epsb$ and $\epsn$. 
We again enforce the constraint that $\fs{} \gtrsim \fd{}$ and $\epsb \gtrsim \epsn$.

Within $6$ months of observation at design sensitivity, we can confirm if the beaming fraction of short gamma-ray bursts launched in NSBH and BNS are different and in what way. 
In particular, with $6$ months of observation for this simulation, we can constrain $\epsb = {0.02}_{-0.02}^{+0.02}$ and $\epsn = {0.16}_{-0.12}^{+0.11}$ (90\% credible interval).

These constraints will help shape our understanding of how gamma-ray burst jets get their structure and could be further decomposed into constraints on the population of parameters corresponding to the Gaussian and power-law structured jets. 
This increase in understanding will be critical for estimating the Hubble constant with coincident gravitational-wave and electromagnetic observations in the absence of very long baseline interferometry \citep[][]{Schutz:1986Natur,LIGOScientific:2017adf,Hotokezaka:2019NatAs}.

\subsection{Constraining binary evolution and GRB physics}
\label{subsec:binaryevolution}

All the models investigated in this work predict jet-launching fractions $\fs{} \lesssim 0.6$ and $\fd{} \lesssim 0.1$. 
The former is enforced due to the constraint that a gamma-ray burst jet is not produced unless the remnant mass is greater than $1.2\times\mtov{}$. 
The latter is dependent on two critical factors. 1) Whether there is a mass gap between neutron stars and black holes and 2) the spins of black holes that participate in NSBH mergers. 

Recently, \citet{Fragione:2021ndl} estimated the fraction of NSBH mergers that tidally disrupt the neutron star for several different equations of states and for different black-hole spin models. 
One model in their work, which estimates black hole spins using single star models from the Geneva stellar evolution code~\citep{Eggenberger2008, Ekstrom2012}, predicts the disruption fraction $\fd{} \gtrsim 0.5$ for all different configurations.
The large $\fd{}$ is attributed to the rapidly rotating black holes produced by the Geneva code, a byproduct of inefficient angular momentum transfer in the massive stars. 
Such a high disruption fraction, and large black hole spins, appear to be inconsistent with the gravitational-wave data \citep[][]{LIGOScientific:2021qlt} and the lack of observed electromagnetic radiation from GW200105 and GW200115~\citeg{zhu21_em,Dichiara2021, Anand2021}. 
We note that this may just be a selection effect, as LIGO preferentially sees more massive systems that are less likely to produce electromagnetic radiation. 

To investigate whether such high disruption fractions can be ruled out with the data, we simulate data with $\fs{} = 0.1$, and $\fd{} = 0.8$ with all other parameters kept the same as above. We again use uninformative priors on the jet launching fractions while the beaming fractions are informed by the Gaussian structured jet model.  

Within $2$ years of observation at design sensitivity, we can rule out $\fd{} \lesssim 0.5$ with $95\%$ confidence. If $\fd{} \gtrsim 0.5$ then this has important implications for binary evolution, as it could either rule out a mass gap between neutron stars and black holes, or imply that the spins of black holes in NSBH are typically much larger than  predicted by the tidal spin up model~\citep[][]{Qin2018}. 
The former would provide insights on the formation of neutron stars and black holes in supernovae, whilst the latter probes the efficiency of angular momentum transport in massive stars and the role of tides in binary evolution.

Neutron star-black hole mergers are often invoked to explain peculiar observations of some short gamma-ray bursts such as the extended prompt emission or internal plateaus~\citeg{desai19, Gompertz2020}. 
However, as noted above, both these features are more typically attributed to the spin-down of a nascent neutron star~\citeg{rowlinson13}, an engine that can not be born in a NSBH merger. 
The fraction of short gamma-ray bursts that exhibit such features is $\approx 30\%$ higher than any of our predictions. 

With parameters informed by the suite of population-synthesis results described earlier and the Gaussian structured jet, we will be able to measure $\fs{}$ and $\fd{}$ with $10\%$ precision within $6$ months of observation at design sensitivity. 
This precision may allow us to determine if the contaminated sample of short gamma-ray bursts with internal plateaus and extended emission are consistent with the disruption fraction of NSBH mergers. This would provide evidence for mechanisms such as fall-back accretion onto the black hole for explaining these peculiar observations~\citeg{desai19}.

Under the constraint that $M_{\rm{rem}} \lesssim 1.2\mtov{}$, all population synthesis models analysed in this work predict $\fs{} \lesssim 0.5$. Models motivated by Galactic and extragalactic observations of double neutron stars~\citeg{sarin20, Galaudage2021} predict slightly higher values of $\fs{} \lesssim 0.6$. 

As described in Sec.~\ref{subsec:fs!=1}, we can rule out if $\fs{} \gtrsim 0.8$ with $90\%$ confidence within a year of observations. Such a high value of $\fs{}$ is inconsistent with any of our model predictions and may suggest that either there are significantly more higher mass mergers than those predicted by population synthesis and current observations of gravitational-wave mergers. Or, that the assumption that $M_{\rm{rem}} \gtrsim 1.2\mtov{}$ is required to produce a gamma-ray burst jet is flawed. Both hypotheses have profound implications on our understanding of binary evolution and gamma-ray burst jet launching. 
\section{Conclusion}
\label{sec:conclusion}
The measurements of rate densities from gravitational-wave observations of BNS and NSBH mergers give us an opportunity to link the observations of short gamma-ray bursts to the mergers that made them through Eq.~\ref{eq:one_to_rule_them_all}. 
This equation is dependent on several terms, each of which is rich in binary and gamma-ray burst physics, as described in Sec.~\ref{sec:physics}. 

We estimate these relevant terms using a suite of population synthesis results and different gamma-ray burst jet structures. In particular, all population synthesis simulations predict between $20-60\%$ and $0.5-10\%$ of BNS and NSBH mergers respectively can successfully launch a jet. Given current constraints on the merger rate and assuming the beaming fraction of gamma-ray bursts can be estimated by $f_b = 1 - \cos(\thj{})$ i.e., that gamma-ray burst jets have a top hat structure we measure the opening angle of short gamma-ray bursts to be $\approx 15^\circ$, consistent with previous estimates. Given current measurements, the choice of jet structure does not change our results, with our constraints assuming a Gaussian structured jet model being identical to the analysis with the top hat structure. 

We also measure the fraction of binary neutron star mergers that can launch a jet to be $\fs{}=0.69^{+0.27}_{-0.37}$ ($90\%$ credible interval) and find that $\fs \gtrsim 0.3$ with $95\%$ confidence, ruling out population synthesis models with rapid supernovae and fixed mass transfer efficiency. 
We provide an updated measurement of the BNS and NSBH merger rates of $\currentrbns{}$ and $\currentrnsbh{}$ ($90\%$ credible interval) respectively, which are approximately $16\%$ and $3\%$ more informative than the BNS and NSBH merger rates from the second gravitational-wave catalog~\citep{LIGOScientific:2020kqk}.

We then perform a series of calculations projecting constraints and questions that can be answered as gravitational-wave observatories observe more events and decrease uncertainties on the observed merger rates. We highlight how these improved constraints allow us to probe several key questions in binary evolution and gamma-ray physics. In Sec.~\ref{sec:forecasts}, we show that within $6-12$ months of observations of the LIGO-Virgo-KAGRA network operating at design sensitivity, we can constrain the fraction of BNS and NSBH mergers that launch a jet to $\sim {10\%}$ providing insights into questions such as the nature of the gamma-ray burst central engine, neutron star binary mass distributions and if BNS and NSBH mergers have different gamma-ray burst structures.

The best constraints derived from Eq.~\ref{eq:one_to_rule_them_all} are on \textit{combinations} of parameters. 
In particular, parameters corresponding to jet-launching and beaming fractions are degenerate with each other.
Similarly, it is currently difficult to disentangle the contributions to GRBs from BNS versus NSBH. 
However, we can use two arguments from first principles to help constrain our priors. 
1) BNS mergers are more likely to launch a jet than NSBH mergers, enforcing a constraint that $\fs{} \gtrsim \fd$ seen across all population-synthesis models considered here (Fig.~\ref{fig:jet-launching-fractions}.). 
2) Gamma-ray burst jets launched in BNS mergers are likely more collimated than those launched in NSBH mergers; i.e., $\epsb \lesssim \epsn$.
Given our current uncertainties on binary and gamma-ray burst physics, improvements to measurements of the rate density alone will be insufficient to measure all the terms in Eq.~\ref{eq:one_to_rule_them_all} beyond a certain number of observations, as we will become limited by systematic error due to uncertain binary and gamma-ray burst physics. 

Fortunately, each of the terms in Eq.~\ref{eq:one_to_rule_them_all} can be constrained with independent observations offering an opportunity to break degeneracies. 
For example, the NSBH and BNS merger rates will improve in the near future as we observe more mergers as the second generation gravitational-wave detectors reach design sensitivity~\citep{Abbott2018}. 
These observations will enable population analyses that improve our understanding of the mass and spin distribution of black holes and masses of neutron stars that participate in such mergers~\citeg{LIGOScientific:2020kqk}. 
Meanwhile multi-messenger observations of events such as GW170817~\citeg{radice18}, or X-ray observations of neutron stars in our Galaxy~\citeg{Riley2019} will independently provide tighter constraints on the nuclear equation of state. 
Both types of observations will dramatically decrease our uncertainty on the jet-launching fractions. 

Independently, studies into short gamma-ray burst afterglows---especially ones that are observed off-axis~\citep[e.g.,][]{troja18_150101B, Sarin2021_cdf}---will improve our understanding of gamma-ray burst jet structure and therefore the beaming fractions of gamma-ray bursts. 
This is especially true for gamma-ray bursts that we observe in coincidence with gravitational-wave events~\citep{Biscoveanu2020, Farah2020, Hayes2020}. 
Such observations will also improve our understanding of the prompt emission generation mechanism~\citeg{kumar15}, and potentially the central engine~\citeg{Sarin2021}, both of which serve to decrease our uncertainty in beaming and jet-launching fraction.

Currently, the sensitivity of gravitational-wave detectors limits us to the local Universe, where measurements of the short gamma-ray burst, NSBH and BNS merger rates does not significantly change with redshift. 
However, third generation gravitational-wave detectors will be able to probe the NSBH and BNS merger rate out to distances where these rates change considerably.
Then it will become necessary to rewrite Eq.~\ref{eq:one_to_rule_them_all} with redshift dependence. This eliminates the systematic uncertainty that arises from estimating the rate density of short gamma-ray bursts in the local Universe using a luminosity function~\citep{Wanderman2015}. However, one may then need to consider the redshift evolution of the parameters that dictate each term, which will complicate things significantly.

Given the rapid advances in the growing field of gravitational-wave astronomy and expected accumulation of observations of neutron star binary mergers, the limiting factor to our analysis is going to be determining the beaming fractions of these mergers. 
Kilonovae offer an opportunity to perform similar analyses to those presented here but without needing to consider the effects of beaming as they emit radiation quasi-spherically. 
Current constraints on the rate of kilonovae are too weak to be useful~\citep{Yang2017, Andreoni2021}, but with surveys with the upcoming Vera C.\ Rubin Observatory and further operation of the Zwicky Transient Facility, among other facilities, the rates will undoubtedly become significantly better. 
This may open up an avenue for applying the framework presented here to both kilonovae and short gamma-ray bursts together. 

\section{Acknowledgments}
We are grateful to Moritz H\"ubner and Floor Broekgaarden for their help in simulating data and extracting properties from {\sc{compas}} simulations in~\cite{Broekgaarden_2021}. 
The authors are supported by the Australian Research Council (ARC) Centre of Excellence for Gravitational Wave Discovery OzGrav, through project number CE170100004, ARC Future Fellowship FT160100112, ARC Linkage Grant LE210100002, ARC Discovery Early Career Research Award DE220100241 and ARC Discovery Project DP180103155.
This work was performed on the OzSTAR national facility at Swinburne University of Technology. 
The OzSTAR program receives funding in part from the Astronomy National Collaborative Research Infrastructure Strategy (NCRIS) allocation provided by the Australian Government.

\bibliography{bibliography}

\begin{thebibliography}{96}
\expandafter\ifx\csname natexlab\endcsname\relax\def\natexlab#1{#1}\fi
\expandafter\ifx\csname bibnamefont\endcsname\relax
  \def\bibnamefont#1{#1}\fi
\expandafter\ifx\csname bibfnamefont\endcsname\relax
  \def\bibfnamefont#1{#1}\fi
\expandafter\ifx\csname citenamefont\endcsname\relax
  \def\citenamefont#1{#1}\fi
\expandafter\ifx\csname url\endcsname\relax
  \def\url#1{\texttt{#1}}\fi
\expandafter\ifx\csname urlprefix\endcsname\relax\def\urlprefix{URL }\fi
\providecommand{\bibinfo}[2]{#2}
\providecommand{\eprint}[2][]{\url{#2}}

\bibitem[{\citenamefont{{Eichler} et~al.}(1989)\citenamefont{{Eichler},
  {Livio}, {Piran}, and {Schramm}}}]{Eichler:1989Natur}
\bibinfo{author}{\bibfnamefont{D.}~\bibnamefont{{Eichler}}},
  \bibinfo{author}{\bibfnamefont{M.}~\bibnamefont{{Livio}}},
  \bibinfo{author}{\bibfnamefont{T.}~\bibnamefont{{Piran}}}, \bibnamefont{and}
  \bibinfo{author}{\bibfnamefont{D.~N.} \bibnamefont{{Schramm}}},
  \bibinfo{journal}{\nat} \textbf{\bibinfo{volume}{340}}, \bibinfo{pages}{126}
  (\bibinfo{year}{1989}).

\bibitem[{\citenamefont{{Narayan} et~al.}(1992)\citenamefont{{Narayan},
  {Paczynski}, and {Piran}}}]{Narayan:1992ApJL}
\bibinfo{author}{\bibfnamefont{R.}~\bibnamefont{{Narayan}}},
  \bibinfo{author}{\bibfnamefont{B.}~\bibnamefont{{Paczynski}}},
  \bibnamefont{and} \bibinfo{author}{\bibfnamefont{T.}~\bibnamefont{{Piran}}},
  \bibinfo{journal}{\apjl} \textbf{\bibinfo{volume}{395}}, \bibinfo{pages}{L83}
  (\bibinfo{year}{1992}), \eprint{astro-ph/9204001}.

\bibitem[{\citenamefont{{Mochkovitch} et~al.}(1993)\citenamefont{{Mochkovitch},
  {Hernanz}, {Isern}, and {Martin}}}]{Mochkovitch:1993Natur}
\bibinfo{author}{\bibfnamefont{R.}~\bibnamefont{{Mochkovitch}}},
  \bibinfo{author}{\bibfnamefont{M.}~\bibnamefont{{Hernanz}}},
  \bibinfo{author}{\bibfnamefont{J.}~\bibnamefont{{Isern}}}, \bibnamefont{and}
  \bibinfo{author}{\bibfnamefont{X.}~\bibnamefont{{Martin}}},
  \bibinfo{journal}{\nat} \textbf{\bibinfo{volume}{361}}, \bibinfo{pages}{236}
  (\bibinfo{year}{1993}).

\bibitem[{\citenamefont{Abbott
  et~al.}(2017{\natexlab{a}})}]{LIGOScientific:2017vwq}
\bibinfo{author}{\bibfnamefont{B.~P.} \bibnamefont{Abbott}}
  \bibnamefont{et~al.} (\bibinfo{collaboration}{LIGO Scientific, Virgo}),
  \bibinfo{journal}{Phys. Rev. Lett.} \textbf{\bibinfo{volume}{119}},
  \bibinfo{pages}{161101} (\bibinfo{year}{2017}{\natexlab{a}}),
  \eprint{1710.05832}.

\bibitem[{\citenamefont{Abbott
  et~al.}(2017{\natexlab{b}})}]{LIGOScientific:2017zic}
\bibinfo{author}{\bibfnamefont{B.~P.} \bibnamefont{Abbott}}
  \bibnamefont{et~al.} (\bibinfo{collaboration}{LIGO Scientific, Virgo,
  Fermi-GBM, INTEGRAL}), \bibinfo{journal}{Astrophys. J. Lett.}
  \textbf{\bibinfo{volume}{848}}, \bibinfo{pages}{L13}
  (\bibinfo{year}{2017}{\natexlab{b}}), \eprint{1710.05834}.

\bibitem[{\citenamefont{{Janka} et~al.}(1999)\citenamefont{{Janka}, {Eberl},
  {Ruffert}, and {Fryer}}}]{Janka:1999ApJL}
\bibinfo{author}{\bibfnamefont{H.~T.} \bibnamefont{{Janka}}},
  \bibinfo{author}{\bibfnamefont{T.}~\bibnamefont{{Eberl}}},
  \bibinfo{author}{\bibfnamefont{M.}~\bibnamefont{{Ruffert}}},
  \bibnamefont{and} \bibinfo{author}{\bibfnamefont{C.~L.}
  \bibnamefont{{Fryer}}}, \bibinfo{journal}{\apjl}
  \textbf{\bibinfo{volume}{527}}, \bibinfo{pages}{L39} (\bibinfo{year}{1999}),
  \eprint{astro-ph/9908290}.

\bibitem[{\citenamefont{{Barbieri} et~al.}(2019)\citenamefont{{Barbieri},
  {Salafia}, {Perego}, {Colpi}, and {Ghirlanda}}}]{Barbieri2019}
\bibinfo{author}{\bibfnamefont{C.}~\bibnamefont{{Barbieri}}},
  \bibinfo{author}{\bibfnamefont{O.~S.} \bibnamefont{{Salafia}}},
  \bibinfo{author}{\bibfnamefont{A.}~\bibnamefont{{Perego}}},
  \bibinfo{author}{\bibfnamefont{M.}~\bibnamefont{{Colpi}}}, \bibnamefont{and}
  \bibinfo{author}{\bibfnamefont{G.}~\bibnamefont{{Ghirlanda}}},
  \bibinfo{journal}{\aap} \textbf{\bibinfo{volume}{625}}, \bibinfo{eid}{A152}
  (\bibinfo{year}{2019}), \eprint{1903.04543}.

\bibitem[{\citenamefont{{Troja} et~al.}(2008)\citenamefont{{Troja}, {King},
  {O'Brien}, {Lyons}, and {Cusumano}}}]{Troja2008}
\bibinfo{author}{\bibfnamefont{E.}~\bibnamefont{{Troja}}},
  \bibinfo{author}{\bibfnamefont{A.~R.} \bibnamefont{{King}}},
  \bibinfo{author}{\bibfnamefont{P.~T.} \bibnamefont{{O'Brien}}},
  \bibinfo{author}{\bibfnamefont{N.}~\bibnamefont{{Lyons}}}, \bibnamefont{and}
  \bibinfo{author}{\bibfnamefont{G.}~\bibnamefont{{Cusumano}}},
  \bibinfo{journal}{\mnras} \textbf{\bibinfo{volume}{385}},
  \bibinfo{pages}{L10} (\bibinfo{year}{2008}), \eprint{0711.3034}.

\bibitem[{\citenamefont{{Siellez} et~al.}(2016)\citenamefont{{Siellez}, {Boer},
  {Gendre}, and {Regimbau}}}]{Siellez2016}
\bibinfo{author}{\bibfnamefont{K.}~\bibnamefont{{Siellez}}},
  \bibinfo{author}{\bibfnamefont{M.}~\bibnamefont{{Boer}}},
  \bibinfo{author}{\bibfnamefont{B.}~\bibnamefont{{Gendre}}}, \bibnamefont{and}
  \bibinfo{author}{\bibfnamefont{T.}~\bibnamefont{{Regimbau}}},
  \bibinfo{journal}{arXiv e-prints} \bibinfo{eid}{arXiv:1606.03043}
  (\bibinfo{year}{2016}), \eprint{1606.03043}.

\bibitem[{\citenamefont{{Gompertz} et~al.}(2020)\citenamefont{{Gompertz},
  {Levan}, and {Tanvir}}}]{Gompertz2020}
\bibinfo{author}{\bibfnamefont{B.~P.} \bibnamefont{{Gompertz}}},
  \bibinfo{author}{\bibfnamefont{A.~J.} \bibnamefont{{Levan}}},
  \bibnamefont{and} \bibinfo{author}{\bibfnamefont{N.~R.}
  \bibnamefont{{Tanvir}}}, \bibinfo{journal}{\apj}
  \textbf{\bibinfo{volume}{895}}, \bibinfo{eid}{58} (\bibinfo{year}{2020}),
  \eprint{2001.08706}.

\bibitem[{\citenamefont{{Burns} et~al.}(2021)\citenamefont{{Burns}, {Svinkin},
  {Hurley}, {Wadiasingh}, {Negro}, {Younes}, {Hamburg}, {Ridnaia}, {Cook},
  {Cenko} et~al.}}]{burns21}
\bibinfo{author}{\bibfnamefont{E.}~\bibnamefont{{Burns}}},
  \bibinfo{author}{\bibfnamefont{D.}~\bibnamefont{{Svinkin}}},
  \bibinfo{author}{\bibfnamefont{K.}~\bibnamefont{{Hurley}}},
  \bibinfo{author}{\bibfnamefont{Z.}~\bibnamefont{{Wadiasingh}}},
  \bibinfo{author}{\bibfnamefont{M.}~\bibnamefont{{Negro}}},
  \bibinfo{author}{\bibfnamefont{G.}~\bibnamefont{{Younes}}},
  \bibinfo{author}{\bibfnamefont{R.}~\bibnamefont{{Hamburg}}},
  \bibinfo{author}{\bibfnamefont{A.}~\bibnamefont{{Ridnaia}}},
  \bibinfo{author}{\bibfnamefont{D.}~\bibnamefont{{Cook}}},
  \bibinfo{author}{\bibfnamefont{S.~B.} \bibnamefont{{Cenko}}},
  \bibnamefont{et~al.}, \bibinfo{journal}{\apjl}
  \textbf{\bibinfo{volume}{907}}, \bibinfo{eid}{L28} (\bibinfo{year}{2021}),
  \eprint{2101.05144}.

\bibitem[{\citenamefont{{Abbott} et~al.}(2018)\citenamefont{{Abbott}, {Abbott},
  {Abbott}, {Abernathy}, and et~al.}}]{Abbott2018}
\bibinfo{author}{\bibfnamefont{B.~P.} \bibnamefont{{Abbott}}},
  \bibinfo{author}{\bibfnamefont{R.}~\bibnamefont{{Abbott}}},
  \bibinfo{author}{\bibfnamefont{T.~D.} \bibnamefont{{Abbott}}},
  \bibinfo{author}{\bibfnamefont{M.~R.} \bibnamefont{{Abernathy}}},
  \bibnamefont{and} \bibinfo{author}{\bibnamefont{et~al.}},
  \bibinfo{journal}{Living Reviews in Relativity}
  \textbf{\bibinfo{volume}{21}}, \bibinfo{eid}{3} (\bibinfo{year}{2018}),
  \eprint{1304.0670}.

\bibitem[{\citenamefont{{Coward} et~al.}(2012)\citenamefont{{Coward}, {Howell},
  {Piran}, {Stratta}, {Branchesi}, {Bromberg}, {Gendre}, {Burman}, and
  {Guetta}}}]{Coward2012}
\bibinfo{author}{\bibfnamefont{D.~M.} \bibnamefont{{Coward}}},
  \bibinfo{author}{\bibfnamefont{E.~J.} \bibnamefont{{Howell}}},
  \bibinfo{author}{\bibfnamefont{T.}~\bibnamefont{{Piran}}},
  \bibinfo{author}{\bibfnamefont{G.}~\bibnamefont{{Stratta}}},
  \bibinfo{author}{\bibfnamefont{M.}~\bibnamefont{{Branchesi}}},
  \bibinfo{author}{\bibfnamefont{O.}~\bibnamefont{{Bromberg}}},
  \bibinfo{author}{\bibfnamefont{B.}~\bibnamefont{{Gendre}}},
  \bibinfo{author}{\bibfnamefont{R.~R.} \bibnamefont{{Burman}}},
  \bibnamefont{and} \bibinfo{author}{\bibfnamefont{D.}~\bibnamefont{{Guetta}}},
  \bibinfo{journal}{\mnras} \textbf{\bibinfo{volume}{425}},
  \bibinfo{pages}{2668} (\bibinfo{year}{2012}), \eprint{1202.2179}.

\bibitem[{\citenamefont{{Wanderman} and {Piran}}(2015)}]{Wanderman2015}
\bibinfo{author}{\bibfnamefont{D.}~\bibnamefont{{Wanderman}}} \bibnamefont{and}
  \bibinfo{author}{\bibfnamefont{T.}~\bibnamefont{{Piran}}},
  \bibinfo{journal}{\mnras} \textbf{\bibinfo{volume}{448}},
  \bibinfo{pages}{3026} (\bibinfo{year}{2015}), \eprint{1405.5878}.

\bibitem[{\citenamefont{{Zhang} and {Wang}}(2018)}]{Zhang2018}
\bibinfo{author}{\bibfnamefont{G.~Q.} \bibnamefont{{Zhang}}} \bibnamefont{and}
  \bibinfo{author}{\bibfnamefont{F.~Y.} \bibnamefont{{Wang}}},
  \bibinfo{journal}{\apj} \textbf{\bibinfo{volume}{852}}, \bibinfo{eid}{1}
  (\bibinfo{year}{2018}), \eprint{1711.08206}.

\bibitem[{\citenamefont{Guo et~al.}(2020)\citenamefont{Guo, Wei, and
  Wang}}]{Guo}
\bibinfo{author}{\bibfnamefont{Q.}~\bibnamefont{Guo}},
  \bibinfo{author}{\bibfnamefont{D.}~\bibnamefont{Wei}}, \bibnamefont{and}
  \bibinfo{author}{\bibfnamefont{Y.}~\bibnamefont{Wang}},
  \bibinfo{journal}{Astrophys. J.} \textbf{\bibinfo{volume}{894}},
  \bibinfo{pages}{11} (\bibinfo{year}{2020}).

\bibitem[{\citenamefont{Abbott
  et~al.}(2021{\natexlab{a}})}]{LIGOScientific:2021qlt}
\bibinfo{author}{\bibfnamefont{R.}~\bibnamefont{Abbott}} \bibnamefont{et~al.}
  (\bibinfo{collaboration}{LIGO Scientific, KAGRA, VIRGO}),
  \bibinfo{journal}{Astrophys. J. Lett.} \textbf{\bibinfo{volume}{915}},
  \bibinfo{pages}{L5} (\bibinfo{year}{2021}{\natexlab{a}}),
  \eprint{2106.15163}.

\bibitem[{\citenamefont{Foucart et~al.}(2018)\citenamefont{Foucart, Hinderer,
  and Nissanke}}]{Foucart_2018}
\bibinfo{author}{\bibfnamefont{F.}~\bibnamefont{Foucart}},
  \bibinfo{author}{\bibfnamefont{T.}~\bibnamefont{Hinderer}}, \bibnamefont{and}
  \bibinfo{author}{\bibfnamefont{S.}~\bibnamefont{Nissanke}},
  \bibinfo{journal}{Phys. Rev. D} \textbf{\bibinfo{volume}{98}},
  \bibinfo{pages}{081501} (\bibinfo{year}{2018}),
  \urlprefix\url{https://link.aps.org/doi/10.1103/PhysRevD.98.081501}.

\bibitem[{\citenamefont{Kr\"uger and Foucart}(2020)}]{Kruger_2020}
\bibinfo{author}{\bibfnamefont{C.~J.} \bibnamefont{Kr\"uger}} \bibnamefont{and}
  \bibinfo{author}{\bibfnamefont{F.}~\bibnamefont{Foucart}},
  \bibinfo{journal}{Phys. Rev. D} \textbf{\bibinfo{volume}{101}},
  \bibinfo{pages}{103002} (\bibinfo{year}{2020}),
  \urlprefix\url{https://link.aps.org/doi/10.1103/PhysRevD.101.103002}.

\bibitem[{\citenamefont{{Zhu} et~al.}(2021)\citenamefont{{Zhu}, {Wu}, {Yang},
  {Zhang}, {Yu}, {Gao}, {Cao}, and {Liu}}}]{zhu21_em}
\bibinfo{author}{\bibfnamefont{J.-P.} \bibnamefont{{Zhu}}},
  \bibinfo{author}{\bibfnamefont{S.}~\bibnamefont{{Wu}}},
  \bibinfo{author}{\bibfnamefont{Y.-P.} \bibnamefont{{Yang}}},
  \bibinfo{author}{\bibfnamefont{B.}~\bibnamefont{{Zhang}}},
  \bibinfo{author}{\bibfnamefont{Y.-W.} \bibnamefont{{Yu}}},
  \bibinfo{author}{\bibfnamefont{H.}~\bibnamefont{{Gao}}},
  \bibinfo{author}{\bibfnamefont{Z.}~\bibnamefont{{Cao}}}, \bibnamefont{and}
  \bibinfo{author}{\bibfnamefont{L.-D.} \bibnamefont{{Liu}}},
  \bibinfo{journal}{arXiv e-prints} \bibinfo{eid}{arXiv:2106.15781}
  (\bibinfo{year}{2021}), \eprint{2106.15781}.

\bibitem[{\citenamefont{{Dichiara} et~al.}(2021)\citenamefont{{Dichiara},
  {Becerra}, {Chase}, {Troja}, and et~al.}}]{Dichiara2021}
\bibinfo{author}{\bibfnamefont{S.}~\bibnamefont{{Dichiara}}},
  \bibinfo{author}{\bibfnamefont{R.~L.} \bibnamefont{{Becerra}}},
  \bibinfo{author}{\bibfnamefont{E.~A.} \bibnamefont{{Chase}}},
  \bibinfo{author}{\bibfnamefont{E.}~\bibnamefont{{Troja}}}, \bibnamefont{and}
  \bibinfo{author}{\bibnamefont{et~al.}}, \bibinfo{journal}{arXiv e-prints}
  \bibinfo{eid}{arXiv:2110.12047} (\bibinfo{year}{2021}), \eprint{2110.12047}.

\bibitem[{\citenamefont{{Anand} et~al.}(2021)\citenamefont{{Anand}, {Coughlin},
  {Kasliwal}, {Bulla}, and et~al.}}]{Anand2021}
\bibinfo{author}{\bibfnamefont{S.}~\bibnamefont{{Anand}}},
  \bibinfo{author}{\bibfnamefont{M.~W.} \bibnamefont{{Coughlin}}},
  \bibinfo{author}{\bibfnamefont{M.~M.} \bibnamefont{{Kasliwal}}},
  \bibinfo{author}{\bibfnamefont{M.}~\bibnamefont{{Bulla}}}, \bibnamefont{and}
  \bibinfo{author}{\bibnamefont{et~al.}}, \bibinfo{journal}{Nature Astronomy}
  \textbf{\bibinfo{volume}{5}}, \bibinfo{pages}{46} (\bibinfo{year}{2021}),
  \eprint{2009.07210}.

\bibitem[{\citenamefont{{Schutz}}(1986)}]{Schutz:1986Natur}
\bibinfo{author}{\bibfnamefont{B.~F.} \bibnamefont{{Schutz}}},
  \bibinfo{journal}{\nat} \textbf{\bibinfo{volume}{323}}, \bibinfo{pages}{310}
  (\bibinfo{year}{1986}).

\bibitem[{\citenamefont{{Vitale} and {Chen}}(2018)}]{Vitale:2018PhRvL}
\bibinfo{author}{\bibfnamefont{S.}~\bibnamefont{{Vitale}}} \bibnamefont{and}
  \bibinfo{author}{\bibfnamefont{H.-Y.} \bibnamefont{{Chen}}},
  \bibinfo{journal}{\prl} \textbf{\bibinfo{volume}{121}}, \bibinfo{eid}{021303}
  (\bibinfo{year}{2018}), \eprint{1804.07337}.

\bibitem[{\citenamefont{{Feeney} et~al.}(2021)\citenamefont{{Feeney}, {Peiris},
  {Nissanke}, and {Mortlock}}}]{Feeney:2021PhRvL}
\bibinfo{author}{\bibfnamefont{S.~M.} \bibnamefont{{Feeney}}},
  \bibinfo{author}{\bibfnamefont{H.~V.} \bibnamefont{{Peiris}}},
  \bibinfo{author}{\bibfnamefont{S.~M.} \bibnamefont{{Nissanke}}},
  \bibnamefont{and} \bibinfo{author}{\bibfnamefont{D.~J.}
  \bibnamefont{{Mortlock}}}, \bibinfo{journal}{\prl}
  \textbf{\bibinfo{volume}{126}}, \bibinfo{eid}{171102} (\bibinfo{year}{2021}),
  \eprint{2012.06593}.

\bibitem[{\citenamefont{Broekgaarden and Berger}(2021)}]{Broekgaarden:2021hlu}
\bibinfo{author}{\bibfnamefont{F.~S.} \bibnamefont{Broekgaarden}}
  \bibnamefont{and} \bibinfo{author}{\bibfnamefont{E.}~\bibnamefont{Berger}}
  (\bibinfo{year}{2021}), \eprint{2108.05763}.

\bibitem[{\citenamefont{Belczynski et~al.}(2002)\citenamefont{Belczynski,
  Kalogera, and Bulik}}]{Belczynski_2002}
\bibinfo{author}{\bibfnamefont{K.}~\bibnamefont{Belczynski}},
  \bibinfo{author}{\bibfnamefont{V.}~\bibnamefont{Kalogera}}, \bibnamefont{and}
  \bibinfo{author}{\bibfnamefont{T.}~\bibnamefont{Bulik}},
  \bibinfo{journal}{The Astrophysical Journal} \textbf{\bibinfo{volume}{572}},
  \bibinfo{pages}{407–431} (\bibinfo{year}{2002}), ISSN
  \bibinfo{issn}{1538-4357}, \urlprefix\url{http://dx.doi.org/10.1086/340304}.

\bibitem[{\citenamefont{Belczynski et~al.}(2006)\citenamefont{Belczynski,
  Perna, Bulik, Kalogera, Ivanova, and Lamb}}]{Belczynski_2006}
\bibinfo{author}{\bibfnamefont{K.}~\bibnamefont{Belczynski}},
  \bibinfo{author}{\bibfnamefont{R.}~\bibnamefont{Perna}},
  \bibinfo{author}{\bibfnamefont{T.}~\bibnamefont{Bulik}},
  \bibinfo{author}{\bibfnamefont{V.}~\bibnamefont{Kalogera}},
  \bibinfo{author}{\bibfnamefont{N.}~\bibnamefont{Ivanova}}, \bibnamefont{and}
  \bibinfo{author}{\bibfnamefont{D.~Q.} \bibnamefont{Lamb}},
  \bibinfo{journal}{The Astrophysical Journal} \textbf{\bibinfo{volume}{648}},
  \bibinfo{pages}{1110–1116} (\bibinfo{year}{2006}), ISSN
  \bibinfo{issn}{1538-4357}, \urlprefix\url{http://dx.doi.org/10.1086/505169}.

\bibitem[{\citenamefont{Broekgaarden et~al.}(2021)\citenamefont{Broekgaarden,
  Berger, Neijssel, Vigna-Gómez, Chattopadhyay, Stevenson, Chruslinska,
  Justham, de~Mink, and Mandel}}]{Broekgaarden_2021}
\bibinfo{author}{\bibfnamefont{F.~S.} \bibnamefont{Broekgaarden}},
  \bibinfo{author}{\bibfnamefont{E.}~\bibnamefont{Berger}},
  \bibinfo{author}{\bibfnamefont{C.~J.} \bibnamefont{Neijssel}},
  \bibinfo{author}{\bibfnamefont{A.}~\bibnamefont{Vigna-Gómez}},
  \bibinfo{author}{\bibfnamefont{D.}~\bibnamefont{Chattopadhyay}},
  \bibinfo{author}{\bibfnamefont{S.}~\bibnamefont{Stevenson}},
  \bibinfo{author}{\bibfnamefont{M.}~\bibnamefont{Chruslinska}},
  \bibinfo{author}{\bibfnamefont{S.}~\bibnamefont{Justham}},
  \bibinfo{author}{\bibfnamefont{S.~E.} \bibnamefont{de~Mink}},
  \bibnamefont{and} \bibinfo{author}{\bibfnamefont{I.}~\bibnamefont{Mandel}},
  \emph{\bibinfo{title}{Impact of massive binary star and cosmic evolution on
  gravitational wave observations i: Black hole - neutron star mergers}}
  (\bibinfo{year}{2021}), \eprint{2103.02608}.

\bibitem[{\citenamefont{{Mandel} and {Broekgaarden}}(2021)}]{Mandel2021_review}
\bibinfo{author}{\bibfnamefont{I.}~\bibnamefont{{Mandel}}} \bibnamefont{and}
  \bibinfo{author}{\bibfnamefont{F.~S.} \bibnamefont{{Broekgaarden}}},
  \bibinfo{journal}{arXiv e-prints} \bibinfo{eid}{arXiv:2107.14239}
  (\bibinfo{year}{2021}), \eprint{2107.14239}.

\bibitem[{\citenamefont{Abbott
  et~al.}(2021{\natexlab{b}})}]{LIGOScientific:2020kqk}
\bibinfo{author}{\bibfnamefont{R.}~\bibnamefont{Abbott}} \bibnamefont{et~al.}
  (\bibinfo{collaboration}{LIGO Scientific, Virgo}),
  \bibinfo{journal}{Astrophys. J. Lett.} \textbf{\bibinfo{volume}{913}},
  \bibinfo{pages}{L7} (\bibinfo{year}{2021}{\natexlab{b}}),
  \eprint{2010.14533}.

\bibitem[{\citenamefont{{Abbott} and others.}(2021)}]{gwtc-3_pop}
\bibinfo{author}{\bibfnamefont{R.}~\bibnamefont{{Abbott}}} \bibnamefont{and}
  \bibinfo{author}{\bibnamefont{others.}}, \bibinfo{journal}{arXiv e-prints}
  \bibinfo{eid}{arXiv:2111.03634} (\bibinfo{year}{2021}), \eprint{2111.03634}.

\bibitem[{\citenamefont{{Sarin} and {Lasky}}(2021)}]{Sarin2021}
\bibinfo{author}{\bibfnamefont{N.}~\bibnamefont{{Sarin}}} \bibnamefont{and}
  \bibinfo{author}{\bibfnamefont{P.~D.} \bibnamefont{{Lasky}}},
  \bibinfo{journal}{General Relativity and Gravitation}
  \textbf{\bibinfo{volume}{53}}, \bibinfo{eid}{59} (\bibinfo{year}{2021}),
  \eprint{2012.08172}.

\bibitem[{\citenamefont{{Margalit} et~al.}(2015)\citenamefont{{Margalit},
  {Metzger}, and {Beloborodov}}}]{margalit15_supramassive}
\bibinfo{author}{\bibfnamefont{B.}~\bibnamefont{{Margalit}}},
  \bibinfo{author}{\bibfnamefont{B.~D.} \bibnamefont{{Metzger}}},
  \bibnamefont{and} \bibinfo{author}{\bibfnamefont{A.~M.}
  \bibnamefont{{Beloborodov}}}, \bibinfo{journal}{\prl}
  \textbf{\bibinfo{volume}{115}}, \bibinfo{eid}{171101} (\bibinfo{year}{2015}),
  \eprint{1505.01842}.

\bibitem[{\citenamefont{{Murguia-Berthier}
  et~al.}(2017)\citenamefont{{Murguia-Berthier}, {Ramirez-Ruiz}, {Montes}, {De
  Colle}, {Rezzolla}, {Rosswog}, {Takami}, {Perego}, and {Lee}}}]{murguia17}
\bibinfo{author}{\bibfnamefont{A.}~\bibnamefont{{Murguia-Berthier}}},
  \bibinfo{author}{\bibfnamefont{E.}~\bibnamefont{{Ramirez-Ruiz}}},
  \bibinfo{author}{\bibfnamefont{G.}~\bibnamefont{{Montes}}},
  \bibinfo{author}{\bibfnamefont{F.}~\bibnamefont{{De Colle}}},
  \bibinfo{author}{\bibfnamefont{L.}~\bibnamefont{{Rezzolla}}},
  \bibinfo{author}{\bibfnamefont{S.}~\bibnamefont{{Rosswog}}},
  \bibinfo{author}{\bibfnamefont{K.}~\bibnamefont{{Takami}}},
  \bibinfo{author}{\bibfnamefont{A.}~\bibnamefont{{Perego}}}, \bibnamefont{and}
  \bibinfo{author}{\bibfnamefont{W.~H.} \bibnamefont{{Lee}}},
  \bibinfo{journal}{\apjl} \textbf{\bibinfo{volume}{835}}, \bibinfo{eid}{L34}
  (\bibinfo{year}{2017}), \eprint{1609.04828}.

\bibitem[{\citenamefont{{Ciolfi}}(2018)}]{ciolfi18_sgrb}
\bibinfo{author}{\bibfnamefont{R.}~\bibnamefont{{Ciolfi}}},
  \bibinfo{journal}{International Journal of Modern Physics D}
  \textbf{\bibinfo{volume}{27}}, \bibinfo{eid}{1842004} (\bibinfo{year}{2018}),
  \eprint{1804.03684}.

\bibitem[{\citenamefont{{Rowlinson} et~al.}(2013)\citenamefont{{Rowlinson},
  {O'Brien}, {Metzger}, {Tanvir}, and {Levan}}}]{rowlinson13}
\bibinfo{author}{\bibfnamefont{A.}~\bibnamefont{{Rowlinson}}},
  \bibinfo{author}{\bibfnamefont{P.~T.} \bibnamefont{{O'Brien}}},
  \bibinfo{author}{\bibfnamefont{B.~D.} \bibnamefont{{Metzger}}},
  \bibinfo{author}{\bibfnamefont{N.~R.} \bibnamefont{{Tanvir}}},
  \bibnamefont{and} \bibinfo{author}{\bibfnamefont{A.~J.}
  \bibnamefont{{Levan}}}, \bibinfo{journal}{\mnras}
  \textbf{\bibinfo{volume}{430}}, \bibinfo{pages}{1061} (\bibinfo{year}{2013}),
  \eprint{1301.0629}.

\bibitem[{\citenamefont{{Sarin} et~al.}(2020)\citenamefont{{Sarin}, {Lasky},
  and {Ashton}}}]{sarin20}
\bibinfo{author}{\bibfnamefont{N.}~\bibnamefont{{Sarin}}},
  \bibinfo{author}{\bibfnamefont{P.~D.} \bibnamefont{{Lasky}}},
  \bibnamefont{and} \bibinfo{author}{\bibfnamefont{G.}~\bibnamefont{{Ashton}}},
  \bibinfo{journal}{\prd} \textbf{\bibinfo{volume}{101}}, \bibinfo{eid}{063021}
  (\bibinfo{year}{2020}), \eprint{2001.06102}.

\bibitem[{\citenamefont{{Rhoads}}(2003)}]{rhoads03}
\bibinfo{author}{\bibfnamefont{J.~E.} \bibnamefont{{Rhoads}}},
  \bibinfo{journal}{\apj} \textbf{\bibinfo{volume}{591}}, \bibinfo{pages}{1097}
  (\bibinfo{year}{2003}), \eprint{astro-ph/0301011}.

\bibitem[{\citenamefont{{Lamb} and {Kobayashi}}(2016)}]{lamb_kobayashi16}
\bibinfo{author}{\bibfnamefont{G.~P.} \bibnamefont{{Lamb}}} \bibnamefont{and}
  \bibinfo{author}{\bibfnamefont{S.}~\bibnamefont{{Kobayashi}}},
  \bibinfo{journal}{\apj} \textbf{\bibinfo{volume}{829}}, \bibinfo{eid}{112}
  (\bibinfo{year}{2016}), \eprint{1605.02769}.

\bibitem[{\citenamefont{Foucart}(2020)}]{Foucart_2020}
\bibinfo{author}{\bibfnamefont{F.}~\bibnamefont{Foucart}},
  \bibinfo{journal}{Frontiers in Astronomy and Space Sciences}
  \textbf{\bibinfo{volume}{7}} (\bibinfo{year}{2020}), ISSN
  \bibinfo{issn}{2296-987X},
  \urlprefix\url{http://dx.doi.org/10.3389/fspas.2020.00046}.

\bibitem[{\citenamefont{Lee and Kluzniak}(1999)}]{Lee_1999}
\bibinfo{author}{\bibfnamefont{W.~H.} \bibnamefont{Lee}} \bibnamefont{and}
  \bibinfo{author}{\bibfnamefont{W.}~\bibnamefont{Kluzniak}},
  \bibinfo{journal}{Monthly Notices of the Royal Astronomical Society}
  \textbf{\bibinfo{volume}{308}}, \bibinfo{pages}{780} (\bibinfo{year}{1999}),
  ISSN \bibinfo{issn}{0035-8711},
  \eprint{https://academic.oup.com/mnras/article-pdf/308/3/780/2959248/308-3-780.pdf},
  \urlprefix\url{https://doi.org/10.1046/j.1365-8711.1999.02734.x}.

\bibitem[{\citenamefont{Wiggins and Lai}(2000)}]{Wiggins_2000}
\bibinfo{author}{\bibfnamefont{P.}~\bibnamefont{Wiggins}} \bibnamefont{and}
  \bibinfo{author}{\bibfnamefont{D.}~\bibnamefont{Lai}}, \bibinfo{journal}{The
  Astrophysical Journal} \textbf{\bibinfo{volume}{532}},
  \bibinfo{pages}{530–539} (\bibinfo{year}{2000}), ISSN
  \bibinfo{issn}{1538-4357}, \urlprefix\url{http://dx.doi.org/10.1086/308565}.

\bibitem[{\citenamefont{{Bardeen} et~al.}(1972)\citenamefont{{Bardeen},
  {Press}, and {Teukolsky}}}]{Bardeen_1972}
\bibinfo{author}{\bibfnamefont{J.~M.} \bibnamefont{{Bardeen}}},
  \bibinfo{author}{\bibfnamefont{W.~H.} \bibnamefont{{Press}}},
  \bibnamefont{and} \bibinfo{author}{\bibfnamefont{S.~A.}
  \bibnamefont{{Teukolsky}}}, \bibinfo{journal}{The Astrophysical Journal}
  \textbf{\bibinfo{volume}{178}}, \bibinfo{pages}{347} (\bibinfo{year}{1972}).

\bibitem[{\citenamefont{{Qin} et~al.}(2018)\citenamefont{{Qin}, {Fragos},
  {Meynet}, {Andrews}, {S{\o}rensen}, and {Song}}}]{Qin2018}
\bibinfo{author}{\bibfnamefont{Y.}~\bibnamefont{{Qin}}},
  \bibinfo{author}{\bibfnamefont{T.}~\bibnamefont{{Fragos}}},
  \bibinfo{author}{\bibfnamefont{G.}~\bibnamefont{{Meynet}}},
  \bibinfo{author}{\bibfnamefont{J.}~\bibnamefont{{Andrews}}},
  \bibinfo{author}{\bibfnamefont{M.}~\bibnamefont{{S{\o}rensen}}},
  \bibnamefont{and} \bibinfo{author}{\bibfnamefont{H.~F.}
  \bibnamefont{{Song}}}, \bibinfo{journal}{\aap}
  \textbf{\bibinfo{volume}{616}}, \bibinfo{eid}{A28} (\bibinfo{year}{2018}),
  \eprint{1802.05738}.

\bibitem[{\citenamefont{{Hernandez Vivanco}
  et~al.}(2020)\citenamefont{{Hernandez Vivanco}, {Smith}, {Thrane}, and
  {Lasky}}}]{HernandezVivanco2020}
\bibinfo{author}{\bibfnamefont{F.}~\bibnamefont{{Hernandez Vivanco}}},
  \bibinfo{author}{\bibfnamefont{R.}~\bibnamefont{{Smith}}},
  \bibinfo{author}{\bibfnamefont{E.}~\bibnamefont{{Thrane}}}, \bibnamefont{and}
  \bibinfo{author}{\bibfnamefont{P.~D.} \bibnamefont{{Lasky}}},
  \bibinfo{journal}{\mnras} \textbf{\bibinfo{volume}{499}},
  \bibinfo{pages}{5972} (\bibinfo{year}{2020}), \eprint{2008.05627}.

\bibitem[{\citenamefont{{Raaijmakers} et~al.}(2021)\citenamefont{{Raaijmakers},
  {Greif}, {Hebeler}, {Hinderer}, {Nissanke}, {Schwenk}, {Riley}, {Watts},
  {Lattimer}, and {Ho}}}]{Raaijmakers2021}
\bibinfo{author}{\bibfnamefont{G.}~\bibnamefont{{Raaijmakers}}},
  \bibinfo{author}{\bibfnamefont{S.~K.} \bibnamefont{{Greif}}},
  \bibinfo{author}{\bibfnamefont{K.}~\bibnamefont{{Hebeler}}},
  \bibinfo{author}{\bibfnamefont{T.}~\bibnamefont{{Hinderer}}},
  \bibinfo{author}{\bibfnamefont{S.}~\bibnamefont{{Nissanke}}},
  \bibinfo{author}{\bibfnamefont{A.}~\bibnamefont{{Schwenk}}},
  \bibinfo{author}{\bibfnamefont{T.~E.} \bibnamefont{{Riley}}},
  \bibinfo{author}{\bibfnamefont{A.~L.} \bibnamefont{{Watts}}},
  \bibinfo{author}{\bibfnamefont{J.~M.} \bibnamefont{{Lattimer}}},
  \bibnamefont{and} \bibinfo{author}{\bibfnamefont{W.~C.~G.}
  \bibnamefont{{Ho}}}, \bibinfo{journal}{\apjl} \textbf{\bibinfo{volume}{918}},
  \bibinfo{eid}{L29} (\bibinfo{year}{2021}), \eprint{2105.06981}.

\bibitem[{\citenamefont{{Smartt} et~al.}(2017)\citenamefont{{Smartt}, {Chen},
  {Jerkstrand}, {Coughlin}, and et~al.}}]{Smartt2017}
\bibinfo{author}{\bibfnamefont{S.~J.} \bibnamefont{{Smartt}}},
  \bibinfo{author}{\bibfnamefont{T.~W.} \bibnamefont{{Chen}}},
  \bibinfo{author}{\bibfnamefont{A.}~\bibnamefont{{Jerkstrand}}},
  \bibinfo{author}{\bibfnamefont{M.}~\bibnamefont{{Coughlin}}},
  \bibnamefont{and} \bibinfo{author}{\bibnamefont{et~al.}},
  \bibinfo{journal}{\nat} \textbf{\bibinfo{volume}{551}}, \bibinfo{pages}{75}
  (\bibinfo{year}{2017}), \eprint{1710.05841}.

\bibitem[{\citenamefont{{Zappa} et~al.}(2019)\citenamefont{{Zappa}, {Bernuzzi},
  {Pannarale}, {Mapelli}, and {Giacobbo}}}]{Zappa2019}
\bibinfo{author}{\bibfnamefont{F.}~\bibnamefont{{Zappa}}},
  \bibinfo{author}{\bibfnamefont{S.}~\bibnamefont{{Bernuzzi}}},
  \bibinfo{author}{\bibfnamefont{F.}~\bibnamefont{{Pannarale}}},
  \bibinfo{author}{\bibfnamefont{M.}~\bibnamefont{{Mapelli}}},
  \bibnamefont{and}
  \bibinfo{author}{\bibfnamefont{N.}~\bibnamefont{{Giacobbo}}},
  \bibinfo{journal}{\prl} \textbf{\bibinfo{volume}{123}}, \bibinfo{eid}{041102}
  (\bibinfo{year}{2019}), \eprint{1903.11622}.

\bibitem[{\citenamefont{{Bhattacharya}
  et~al.}(2019)\citenamefont{{Bhattacharya}, {Kumar}, and
  {Smoot}}}]{Bhattacharya2019}
\bibinfo{author}{\bibfnamefont{M.}~\bibnamefont{{Bhattacharya}}},
  \bibinfo{author}{\bibfnamefont{P.}~\bibnamefont{{Kumar}}}, \bibnamefont{and}
  \bibinfo{author}{\bibfnamefont{G.}~\bibnamefont{{Smoot}}},
  \bibinfo{journal}{\mnras} \textbf{\bibinfo{volume}{486}},
  \bibinfo{pages}{5289} (\bibinfo{year}{2019}), \eprint{1809.00006}.

\bibitem[{\citenamefont{{Galaudage} et~al.}(2021)\citenamefont{{Galaudage},
  {Adamcewicz}, {Zhu}, {Stevenson}, and {Thrane}}}]{Galaudage2021}
\bibinfo{author}{\bibfnamefont{S.}~\bibnamefont{{Galaudage}}},
  \bibinfo{author}{\bibfnamefont{C.}~\bibnamefont{{Adamcewicz}}},
  \bibinfo{author}{\bibfnamefont{X.-J.} \bibnamefont{{Zhu}}},
  \bibinfo{author}{\bibfnamefont{S.}~\bibnamefont{{Stevenson}}},
  \bibnamefont{and} \bibinfo{author}{\bibfnamefont{E.}~\bibnamefont{{Thrane}}},
  \bibinfo{journal}{\apjl} \textbf{\bibinfo{volume}{909}}, \bibinfo{eid}{L19}
  (\bibinfo{year}{2021}), \eprint{2011.01495}.

\bibitem[{\citenamefont{{Fryer} et~al.}(2012)\citenamefont{{Fryer},
  {Belczynski}, {Wiktorowicz}, {Dominik}, {Kalogera}, and {Holz}}}]{Fryer2012}
\bibinfo{author}{\bibfnamefont{C.~L.} \bibnamefont{{Fryer}}},
  \bibinfo{author}{\bibfnamefont{K.}~\bibnamefont{{Belczynski}}},
  \bibinfo{author}{\bibfnamefont{G.}~\bibnamefont{{Wiktorowicz}}},
  \bibinfo{author}{\bibfnamefont{M.}~\bibnamefont{{Dominik}}},
  \bibinfo{author}{\bibfnamefont{V.}~\bibnamefont{{Kalogera}}},
  \bibnamefont{and} \bibinfo{author}{\bibfnamefont{D.~E.}
  \bibnamefont{{Holz}}}, \bibinfo{journal}{\apj}
  \textbf{\bibinfo{volume}{749}}, \bibinfo{eid}{91} (\bibinfo{year}{2012}),
  \eprint{1110.1726}.

\bibitem[{\citenamefont{{{\"O}zel} et~al.}(2010)\citenamefont{{{\"O}zel},
  {Psaltis}, {Narayan}, and {McClintock}}}]{Ozel:2010ApJ}
\bibinfo{author}{\bibfnamefont{F.}~\bibnamefont{{{\"O}zel}}},
  \bibinfo{author}{\bibfnamefont{D.}~\bibnamefont{{Psaltis}}},
  \bibinfo{author}{\bibfnamefont{R.}~\bibnamefont{{Narayan}}},
  \bibnamefont{and} \bibinfo{author}{\bibfnamefont{J.~E.}
  \bibnamefont{{McClintock}}}, \bibinfo{journal}{\apj}
  \textbf{\bibinfo{volume}{725}}, \bibinfo{pages}{1918} (\bibinfo{year}{2010}),
  \eprint{1006.2834}.

\bibitem[{\citenamefont{{Farr} et~al.}(2011)\citenamefont{{Farr}, {Sravan},
  {Cantrell}, {Kreidberg}, {Bailyn}, {Mandel}, and {Kalogera}}}]{Farr:2011ApJ}
\bibinfo{author}{\bibfnamefont{W.~M.} \bibnamefont{{Farr}}},
  \bibinfo{author}{\bibfnamefont{N.}~\bibnamefont{{Sravan}}},
  \bibinfo{author}{\bibfnamefont{A.}~\bibnamefont{{Cantrell}}},
  \bibinfo{author}{\bibfnamefont{L.}~\bibnamefont{{Kreidberg}}},
  \bibinfo{author}{\bibfnamefont{C.~D.} \bibnamefont{{Bailyn}}},
  \bibinfo{author}{\bibfnamefont{I.}~\bibnamefont{{Mandel}}}, \bibnamefont{and}
  \bibinfo{author}{\bibfnamefont{V.}~\bibnamefont{{Kalogera}}},
  \bibinfo{journal}{\apj} \textbf{\bibinfo{volume}{741}}, \bibinfo{eid}{103}
  (\bibinfo{year}{2011}), \eprint{1011.1459}.

\bibitem[{\citenamefont{Chattopadhyay et~al.}(2021)\citenamefont{Chattopadhyay,
  Stevenson, Hurley, Bailes, and Broekgaarden}}]{Chattopadhyay_2021}
\bibinfo{author}{\bibfnamefont{D.}~\bibnamefont{Chattopadhyay}},
  \bibinfo{author}{\bibfnamefont{S.}~\bibnamefont{Stevenson}},
  \bibinfo{author}{\bibfnamefont{J.~R.} \bibnamefont{Hurley}},
  \bibinfo{author}{\bibfnamefont{M.}~\bibnamefont{Bailes}}, \bibnamefont{and}
  \bibinfo{author}{\bibfnamefont{F.}~\bibnamefont{Broekgaarden}},
  \bibinfo{journal}{Monthly Notices of the Royal Astronomical Society}
  \textbf{\bibinfo{volume}{504}}, \bibinfo{pages}{3682–3710}
  (\bibinfo{year}{2021}), ISSN \bibinfo{issn}{1365-2966},
  \urlprefix\url{http://dx.doi.org/10.1093/mnras/stab973}.

\bibitem[{\citenamefont{{Alsing} et~al.}(2018)\citenamefont{{Alsing}, {Silva},
  and {Berti}}}]{alsing18}
\bibinfo{author}{\bibfnamefont{J.}~\bibnamefont{{Alsing}}},
  \bibinfo{author}{\bibfnamefont{H.~O.} \bibnamefont{{Silva}}},
  \bibnamefont{and} \bibinfo{author}{\bibfnamefont{E.}~\bibnamefont{{Berti}}},
  \bibinfo{journal}{\mnras} \textbf{\bibinfo{volume}{478}},
  \bibinfo{pages}{1377} (\bibinfo{year}{2018}), \eprint{1709.07889}.

\bibitem[{\citenamefont{{Margalit} and {Metzger}}(2019)}]{margalit19}
\bibinfo{author}{\bibfnamefont{B.}~\bibnamefont{{Margalit}}} \bibnamefont{and}
  \bibinfo{author}{\bibfnamefont{B.~D.} \bibnamefont{{Metzger}}},
  \bibinfo{journal}{\apjl} \textbf{\bibinfo{volume}{880}}, \bibinfo{eid}{L15}
  (\bibinfo{year}{2019}), \eprint{1904.11995}.

\bibitem[{\citenamefont{{Beniamini} et~al.}(2019)\citenamefont{{Beniamini},
  {Petropoulou}, {Barniol Duran}, and {Giannios}}}]{Beniamini2019}
\bibinfo{author}{\bibfnamefont{P.}~\bibnamefont{{Beniamini}}},
  \bibinfo{author}{\bibfnamefont{M.}~\bibnamefont{{Petropoulou}}},
  \bibinfo{author}{\bibfnamefont{R.}~\bibnamefont{{Barniol Duran}}},
  \bibnamefont{and}
  \bibinfo{author}{\bibfnamefont{D.}~\bibnamefont{{Giannios}}},
  \bibinfo{journal}{\mnras} \textbf{\bibinfo{volume}{483}},
  \bibinfo{pages}{840} (\bibinfo{year}{2019}), \eprint{1808.04831}.

\bibitem[{\citenamefont{Vigna-G\'omez et~al.}(2018)}]{Vigna-Gomez:2018dza}
\bibinfo{author}{\bibfnamefont{A.}~\bibnamefont{Vigna-G\'omez}}
  \bibnamefont{et~al.}, \bibinfo{journal}{Mon. Not. Roy. Astron. Soc.}
  \textbf{\bibinfo{volume}{481}}, \bibinfo{pages}{4009} (\bibinfo{year}{2018}),
  \eprint{1805.07974}.

\bibitem[{\citenamefont{Chattopadhyay et~al.}(2020)\citenamefont{Chattopadhyay,
  Stevenson, Hurley, Rossi, and Flynn}}]{Chattopadhyay:2019xye}
\bibinfo{author}{\bibfnamefont{D.}~\bibnamefont{Chattopadhyay}},
  \bibinfo{author}{\bibfnamefont{S.}~\bibnamefont{Stevenson}},
  \bibinfo{author}{\bibfnamefont{J.~R.} \bibnamefont{Hurley}},
  \bibinfo{author}{\bibfnamefont{L.~J.} \bibnamefont{Rossi}}, \bibnamefont{and}
  \bibinfo{author}{\bibfnamefont{C.}~\bibnamefont{Flynn}},
  \bibinfo{journal}{Mon. Not. Roy. Astron. Soc.}
  \textbf{\bibinfo{volume}{494}}, \bibinfo{pages}{1587} (\bibinfo{year}{2020}),
  \eprint{1912.02415}.

\bibitem[{\citenamefont{Mandel and Smith}(2021)}]{Mandel:2021ewy}
\bibinfo{author}{\bibfnamefont{I.}~\bibnamefont{Mandel}} \bibnamefont{and}
  \bibinfo{author}{\bibfnamefont{R.~J.~E.} \bibnamefont{Smith}}
  (\bibinfo{year}{2021}), \eprint{2109.14759}.

\bibitem[{\citenamefont{Abbott et~al.}(2020)}]{LIGOScientific:2020aai}
\bibinfo{author}{\bibfnamefont{B.~P.} \bibnamefont{Abbott}}
  \bibnamefont{et~al.} (\bibinfo{collaboration}{LIGO Scientific, Virgo}),
  \bibinfo{journal}{Astrophys. J. Lett.} \textbf{\bibinfo{volume}{892}},
  \bibinfo{pages}{L3} (\bibinfo{year}{2020}), \eprint{2001.01761}.

\bibitem[{\citenamefont{Vigna-G\'omez et~al.}(2021)\citenamefont{Vigna-G\'omez,
  Schr\o{}der, Ramirez-Ruiz, Aguilera-Dena, Batta, Langer, and
  Willcox}}]{Vigna-Gomez:2021oqy}
\bibinfo{author}{\bibfnamefont{A.}~\bibnamefont{Vigna-G\'omez}},
  \bibinfo{author}{\bibfnamefont{S.~L.} \bibnamefont{Schr\o{}der}},
  \bibinfo{author}{\bibfnamefont{E.}~\bibnamefont{Ramirez-Ruiz}},
  \bibinfo{author}{\bibfnamefont{D.~R.} \bibnamefont{Aguilera-Dena}},
  \bibinfo{author}{\bibfnamefont{A.}~\bibnamefont{Batta}},
  \bibinfo{author}{\bibfnamefont{N.}~\bibnamefont{Langer}}, \bibnamefont{and}
  \bibinfo{author}{\bibfnamefont{R.}~\bibnamefont{Willcox}},
  \bibinfo{journal}{Astrophys. J. Lett.} \textbf{\bibinfo{volume}{920}},
  \bibinfo{pages}{L17} (\bibinfo{year}{2021}), \eprint{2106.12381}.

\bibitem[{\citenamefont{{Kiziltan} et~al.}(2013)\citenamefont{{Kiziltan},
  {Kottas}, {De Yoreo}, and {Thorsett}}}]{kiziltan13}
\bibinfo{author}{\bibfnamefont{B.}~\bibnamefont{{Kiziltan}}},
  \bibinfo{author}{\bibfnamefont{A.}~\bibnamefont{{Kottas}}},
  \bibinfo{author}{\bibfnamefont{M.}~\bibnamefont{{De Yoreo}}},
  \bibnamefont{and} \bibinfo{author}{\bibfnamefont{S.~E.}
  \bibnamefont{{Thorsett}}}, \bibinfo{journal}{\apj}
  \textbf{\bibinfo{volume}{778}}, \bibinfo{eid}{66} (\bibinfo{year}{2013}),
  \eprint{1011.4291}.

\bibitem[{\citenamefont{{Farrow} et~al.}(2019)\citenamefont{{Farrow}, {Zhu},
  and {Thrane}}}]{farrow19}
\bibinfo{author}{\bibfnamefont{N.}~\bibnamefont{{Farrow}}},
  \bibinfo{author}{\bibfnamefont{X.-J.} \bibnamefont{{Zhu}}}, \bibnamefont{and}
  \bibinfo{author}{\bibfnamefont{E.}~\bibnamefont{{Thrane}}},
  \bibinfo{journal}{\apj} \textbf{\bibinfo{volume}{876}}, \bibinfo{eid}{18}
  (\bibinfo{year}{2019}), \eprint{1902.03300}.

\bibitem[{\citenamefont{{Frail} et~al.}(2001)\citenamefont{{Frail}, {Kulkarni},
  {Sari}, {Djorgovski}, and et~al.}}]{Frail2001}
\bibinfo{author}{\bibfnamefont{D.~A.} \bibnamefont{{Frail}}},
  \bibinfo{author}{\bibfnamefont{S.~R.} \bibnamefont{{Kulkarni}}},
  \bibinfo{author}{\bibfnamefont{R.}~\bibnamefont{{Sari}}},
  \bibinfo{author}{\bibfnamefont{S.~G.} \bibnamefont{{Djorgovski}}},
  \bibnamefont{and} \bibinfo{author}{\bibnamefont{et~al.}},
  \bibinfo{journal}{\apjl} \textbf{\bibinfo{volume}{562}}, \bibinfo{pages}{L55}
  (\bibinfo{year}{2001}), \eprint{astro-ph/0102282}.

\bibitem[{\citenamefont{{Berger}}(2014)}]{berger14}
\bibinfo{author}{\bibfnamefont{E.}~\bibnamefont{{Berger}}},
  \bibinfo{journal}{\araa} \textbf{\bibinfo{volume}{52}}, \bibinfo{pages}{43}
  (\bibinfo{year}{2014}), \eprint{1311.2603}.

\bibitem[{\citenamefont{{Kyutoku} et~al.}(2021)\citenamefont{{Kyutoku},
  {Shibata}, and {Taniguchi}}}]{Kyutoku2021}
\bibinfo{author}{\bibfnamefont{K.}~\bibnamefont{{Kyutoku}}},
  \bibinfo{author}{\bibfnamefont{M.}~\bibnamefont{{Shibata}}},
  \bibnamefont{and}
  \bibinfo{author}{\bibfnamefont{K.}~\bibnamefont{{Taniguchi}}},
  \bibinfo{journal}{arXiv e-prints} \bibinfo{eid}{arXiv:2110.06218}
  (\bibinfo{year}{2021}), \eprint{2110.06218}.

\bibitem[{\citenamefont{{Urrutia} et~al.}(2021)\citenamefont{{Urrutia}, {De
  Colle}, {Murguia-Berthier}, and {Ramirez-Ruiz}}}]{Urrutia2021}
\bibinfo{author}{\bibfnamefont{G.}~\bibnamefont{{Urrutia}}},
  \bibinfo{author}{\bibfnamefont{F.}~\bibnamefont{{De Colle}}},
  \bibinfo{author}{\bibfnamefont{A.}~\bibnamefont{{Murguia-Berthier}}},
  \bibnamefont{and}
  \bibinfo{author}{\bibfnamefont{E.}~\bibnamefont{{Ramirez-Ruiz}}},
  \bibinfo{journal}{\mnras} \textbf{\bibinfo{volume}{503}},
  \bibinfo{pages}{4363} (\bibinfo{year}{2021}), \eprint{2011.06729}.

\bibitem[{\citenamefont{{Just} et~al.}(2016)\citenamefont{{Just},
  {Obergaulinger}, {Janka}, {Bauswein}, and {Schwarz}}}]{Just2016}
\bibinfo{author}{\bibfnamefont{O.}~\bibnamefont{{Just}}},
  \bibinfo{author}{\bibfnamefont{M.}~\bibnamefont{{Obergaulinger}}},
  \bibinfo{author}{\bibfnamefont{H.~T.} \bibnamefont{{Janka}}},
  \bibinfo{author}{\bibfnamefont{A.}~\bibnamefont{{Bauswein}}},
  \bibnamefont{and}
  \bibinfo{author}{\bibfnamefont{N.}~\bibnamefont{{Schwarz}}},
  \bibinfo{journal}{\apjl} \textbf{\bibinfo{volume}{816}}, \bibinfo{eid}{L30}
  (\bibinfo{year}{2016}), \eprint{1510.04288}.

\bibitem[{\citenamefont{{Matsumoto} et~al.}(2019)\citenamefont{{Matsumoto},
  {Nakar}, and {Piran}}}]{matsumoto19}
\bibinfo{author}{\bibfnamefont{T.}~\bibnamefont{{Matsumoto}}},
  \bibinfo{author}{\bibfnamefont{E.}~\bibnamefont{{Nakar}}}, \bibnamefont{and}
  \bibinfo{author}{\bibfnamefont{T.}~\bibnamefont{{Piran}}},
  \bibinfo{journal}{\mnras} \textbf{\bibinfo{volume}{483}},
  \bibinfo{pages}{1247} (\bibinfo{year}{2019}), \eprint{1807.04756}.

\bibitem[{\citenamefont{{Troja} et~al.}(2018)\citenamefont{{Troja}, {Ryan},
  {Piro}, {van Eerten}, {Cenko}, {Yoon}, {Lee}, {Im}, {Sakamoto}, {Gatkine}
  et~al.}}]{troja18_150101B}
\bibinfo{author}{\bibfnamefont{E.}~\bibnamefont{{Troja}}},
  \bibinfo{author}{\bibfnamefont{G.}~\bibnamefont{{Ryan}}},
  \bibinfo{author}{\bibfnamefont{L.}~\bibnamefont{{Piro}}},
  \bibinfo{author}{\bibfnamefont{H.}~\bibnamefont{{van Eerten}}},
  \bibinfo{author}{\bibfnamefont{S.~B.} \bibnamefont{{Cenko}}},
  \bibinfo{author}{\bibfnamefont{Y.}~\bibnamefont{{Yoon}}},
  \bibinfo{author}{\bibfnamefont{S.~K.} \bibnamefont{{Lee}}},
  \bibinfo{author}{\bibfnamefont{M.}~\bibnamefont{{Im}}},
  \bibinfo{author}{\bibfnamefont{T.}~\bibnamefont{{Sakamoto}}},
  \bibinfo{author}{\bibfnamefont{P.}~\bibnamefont{{Gatkine}}},
  \bibnamefont{et~al.}, \bibinfo{journal}{Nature Communications}
  \textbf{\bibinfo{volume}{9}}, \bibinfo{eid}{4089} (\bibinfo{year}{2018}).

\bibitem[{\citenamefont{{Howell} et~al.}(2019)\citenamefont{{Howell}, {Ackley},
  {Rowlinson}, and {Coward}}}]{howell19}
\bibinfo{author}{\bibfnamefont{E.~J.} \bibnamefont{{Howell}}},
  \bibinfo{author}{\bibfnamefont{K.}~\bibnamefont{{Ackley}}},
  \bibinfo{author}{\bibfnamefont{A.}~\bibnamefont{{Rowlinson}}},
  \bibnamefont{and} \bibinfo{author}{\bibfnamefont{D.}~\bibnamefont{{Coward}}},
  \bibinfo{journal}{\mnras} \textbf{\bibinfo{volume}{485}},
  \bibinfo{pages}{1435} (\bibinfo{year}{2019}), \eprint{1811.09168}.

\bibitem[{\citenamefont{{Salafia} et~al.}(2019)\citenamefont{{Salafia},
  {Ghirlanda}, {Ascenzi}, and {Ghisellini}}}]{Salafia2019}
\bibinfo{author}{\bibfnamefont{O.~S.} \bibnamefont{{Salafia}}},
  \bibinfo{author}{\bibfnamefont{G.}~\bibnamefont{{Ghirlanda}}},
  \bibinfo{author}{\bibfnamefont{S.}~\bibnamefont{{Ascenzi}}},
  \bibnamefont{and}
  \bibinfo{author}{\bibfnamefont{G.}~\bibnamefont{{Ghisellini}}},
  \bibinfo{journal}{\aap} \textbf{\bibinfo{volume}{628}}, \bibinfo{eid}{A18}
  (\bibinfo{year}{2019}), \eprint{1905.01190}.

\bibitem[{\citenamefont{{Cunningham} et~al.}(2020)\citenamefont{{Cunningham},
  {Cenko}, {Ryan}, {Vogel}, and et~al.}}]{Cunningham2020}
\bibinfo{author}{\bibfnamefont{V.}~\bibnamefont{{Cunningham}}},
  \bibinfo{author}{\bibfnamefont{S.~B.} \bibnamefont{{Cenko}}},
  \bibinfo{author}{\bibfnamefont{G.}~\bibnamefont{{Ryan}}},
  \bibinfo{author}{\bibfnamefont{S.~N.} \bibnamefont{{Vogel}}},
  \bibnamefont{and} \bibinfo{author}{\bibnamefont{et~al.}},
  \bibinfo{journal}{\apj} \textbf{\bibinfo{volume}{904}}, \bibinfo{eid}{166}
  (\bibinfo{year}{2020}), \eprint{2009.00579}.

\bibitem[{\citenamefont{{Ghirlanda} et~al.}(2019)\citenamefont{{Ghirlanda},
  {Salafia}, {Paragi}, {Giroletti}, and et~al.}}]{Ghirlanda2019}
\bibinfo{author}{\bibfnamefont{G.}~\bibnamefont{{Ghirlanda}}},
  \bibinfo{author}{\bibfnamefont{O.~S.} \bibnamefont{{Salafia}}},
  \bibinfo{author}{\bibfnamefont{Z.}~\bibnamefont{{Paragi}}},
  \bibinfo{author}{\bibfnamefont{M.}~\bibnamefont{{Giroletti}}},
  \bibnamefont{and} \bibinfo{author}{\bibnamefont{et~al.}},
  \bibinfo{journal}{Science} \textbf{\bibinfo{volume}{363}},
  \bibinfo{pages}{968} (\bibinfo{year}{2019}), \eprint{1808.00469}.

\bibitem[{\citenamefont{{Biscoveanu} et~al.}(2020)\citenamefont{{Biscoveanu},
  {Thrane}, and {Vitale}}}]{Biscoveanu2020}
\bibinfo{author}{\bibfnamefont{S.}~\bibnamefont{{Biscoveanu}}},
  \bibinfo{author}{\bibfnamefont{E.}~\bibnamefont{{Thrane}}}, \bibnamefont{and}
  \bibinfo{author}{\bibfnamefont{S.}~\bibnamefont{{Vitale}}},
  \bibinfo{journal}{\apj} \textbf{\bibinfo{volume}{893}}, \bibinfo{eid}{38}
  (\bibinfo{year}{2020}), \eprint{1911.01379}.

\bibitem[{\citenamefont{{Farah} et~al.}(2020)\citenamefont{{Farah}, {Essick},
  {Doctor}, {Fishbach}, and {Holz}}}]{Farah2020}
\bibinfo{author}{\bibfnamefont{A.}~\bibnamefont{{Farah}}},
  \bibinfo{author}{\bibfnamefont{R.}~\bibnamefont{{Essick}}},
  \bibinfo{author}{\bibfnamefont{Z.}~\bibnamefont{{Doctor}}},
  \bibinfo{author}{\bibfnamefont{M.}~\bibnamefont{{Fishbach}}},
  \bibnamefont{and} \bibinfo{author}{\bibfnamefont{D.~E.}
  \bibnamefont{{Holz}}}, \bibinfo{journal}{\apj}
  \textbf{\bibinfo{volume}{895}}, \bibinfo{eid}{108} (\bibinfo{year}{2020}),
  \eprint{1912.04906}.

\bibitem[{\citenamefont{{Williams} et~al.}(2018)\citenamefont{{Williams},
  {Clark}, {Williamson}, and {Heng}}}]{Williams2018}
\bibinfo{author}{\bibfnamefont{D.}~\bibnamefont{{Williams}}},
  \bibinfo{author}{\bibfnamefont{J.~A.} \bibnamefont{{Clark}}},
  \bibinfo{author}{\bibfnamefont{A.~R.} \bibnamefont{{Williamson}}},
  \bibnamefont{and} \bibinfo{author}{\bibfnamefont{I.~S.}
  \bibnamefont{{Heng}}}, \bibinfo{journal}{\apj}
  \textbf{\bibinfo{volume}{858}}, \bibinfo{eid}{79} (\bibinfo{year}{2018}),
  \eprint{1712.02585}.

\bibitem[{\citenamefont{{Fong} et~al.}(2012)\citenamefont{{Fong}, {Berger},
  {Margutti}, {Zauderer}, and et~al.}}]{Fong2012}
\bibinfo{author}{\bibfnamefont{W.}~\bibnamefont{{Fong}}},
  \bibinfo{author}{\bibfnamefont{E.}~\bibnamefont{{Berger}}},
  \bibinfo{author}{\bibfnamefont{R.}~\bibnamefont{{Margutti}}},
  \bibinfo{author}{\bibfnamefont{B.~A.} \bibnamefont{{Zauderer}}},
  \bibnamefont{and} \bibinfo{author}{\bibnamefont{et~al.}},
  \bibinfo{journal}{\apj} \textbf{\bibinfo{volume}{756}}, \bibinfo{eid}{189}
  (\bibinfo{year}{2012}), \eprint{1204.5475}.

\bibitem[{\citenamefont{{Nativi} et~al.}(2021)\citenamefont{{Nativi}, {Lamb},
  {Rosswog}, {Lundman}, and {Kowal}}}]{Nativi2021}
\bibinfo{author}{\bibfnamefont{L.}~\bibnamefont{{Nativi}}},
  \bibinfo{author}{\bibfnamefont{G.~P.} \bibnamefont{{Lamb}}},
  \bibinfo{author}{\bibfnamefont{S.}~\bibnamefont{{Rosswog}}},
  \bibinfo{author}{\bibfnamefont{C.}~\bibnamefont{{Lundman}}},
  \bibnamefont{and} \bibinfo{author}{\bibfnamefont{G.}~\bibnamefont{{Kowal}}},
  \bibinfo{journal}{arXiv e-prints} \bibinfo{eid}{arXiv:2109.00814}
  (\bibinfo{year}{2021}), \eprint{2109.00814}.

\bibitem[{\citenamefont{{Gehrels} et~al.}(2004)\citenamefont{{Gehrels},
  {Chincarini}, {Giommi}, {Mason}, and et~al.}}]{swift}
\bibinfo{author}{\bibfnamefont{N.}~\bibnamefont{{Gehrels}}},
  \bibinfo{author}{\bibfnamefont{G.}~\bibnamefont{{Chincarini}}},
  \bibinfo{author}{\bibfnamefont{P.}~\bibnamefont{{Giommi}}},
  \bibinfo{author}{\bibfnamefont{K.~O.} \bibnamefont{{Mason}}},
  \bibnamefont{and} \bibinfo{author}{\bibnamefont{et~al.}},
  \bibinfo{journal}{\apj} \textbf{\bibinfo{volume}{611}}, \bibinfo{pages}{1005}
  (\bibinfo{year}{2004}), \eprint{astro-ph/0405233}.

\bibitem[{\citenamefont{Chatziioannou and Farr}(2020)}]{Chatziioannou:2020msi}
\bibinfo{author}{\bibfnamefont{K.}~\bibnamefont{Chatziioannou}}
  \bibnamefont{and} \bibinfo{author}{\bibfnamefont{W.~M.} \bibnamefont{Farr}},
  \bibinfo{journal}{Phys. Rev. D} \textbf{\bibinfo{volume}{102}},
  \bibinfo{pages}{064063} (\bibinfo{year}{2020}), \eprint{2005.00482}.

\bibitem[{\citenamefont{Abbott
  et~al.}(2017{\natexlab{c}})}]{LIGOScientific:2017adf}
\bibinfo{author}{\bibfnamefont{B.~P.} \bibnamefont{Abbott}}
  \bibnamefont{et~al.} (\bibinfo{collaboration}{LIGO Scientific, Virgo, 1M2H,
  Dark Energy Camera GW-E, DES, DLT40, Las Cumbres Observatory, VINROUGE,
  MASTER}), \bibinfo{journal}{Nature} \textbf{\bibinfo{volume}{551}},
  \bibinfo{pages}{85} (\bibinfo{year}{2017}{\natexlab{c}}),
  \eprint{1710.05835}.

\bibitem[{\citenamefont{{Hotokezaka} et~al.}(2019)\citenamefont{{Hotokezaka},
  {Nakar}, {Gottlieb}, {Nissanke}, {Masuda}, {Hallinan}, {Mooley}, and
  {Deller}}}]{Hotokezaka:2019NatAs}
\bibinfo{author}{\bibfnamefont{K.}~\bibnamefont{{Hotokezaka}}},
  \bibinfo{author}{\bibfnamefont{E.}~\bibnamefont{{Nakar}}},
  \bibinfo{author}{\bibfnamefont{O.}~\bibnamefont{{Gottlieb}}},
  \bibinfo{author}{\bibfnamefont{S.}~\bibnamefont{{Nissanke}}},
  \bibinfo{author}{\bibfnamefont{K.}~\bibnamefont{{Masuda}}},
  \bibinfo{author}{\bibfnamefont{G.}~\bibnamefont{{Hallinan}}},
  \bibinfo{author}{\bibfnamefont{K.~P.} \bibnamefont{{Mooley}}},
  \bibnamefont{and} \bibinfo{author}{\bibfnamefont{A.~T.}
  \bibnamefont{{Deller}}}, \bibinfo{journal}{Nature Astronomy}
  \textbf{\bibinfo{volume}{3}}, \bibinfo{pages}{940} (\bibinfo{year}{2019}),
  \eprint{1806.10596}.

\bibitem[{\citenamefont{Fragione}(2021)}]{Fragione:2021ndl}
\bibinfo{author}{\bibfnamefont{G.}~\bibnamefont{Fragione}}
  (\bibinfo{year}{2021}), \eprint{2110.09604}.

\bibitem[{\citenamefont{{Eggenberger} et~al.}(2008)\citenamefont{{Eggenberger},
  {Meynet}, {Maeder}, {Hirschi}, {Charbonnel}, {Talon}, and
  {Ekstr{\"o}m}}}]{Eggenberger2008}
\bibinfo{author}{\bibfnamefont{P.}~\bibnamefont{{Eggenberger}}},
  \bibinfo{author}{\bibfnamefont{G.}~\bibnamefont{{Meynet}}},
  \bibinfo{author}{\bibfnamefont{A.}~\bibnamefont{{Maeder}}},
  \bibinfo{author}{\bibfnamefont{R.}~\bibnamefont{{Hirschi}}},
  \bibinfo{author}{\bibfnamefont{C.}~\bibnamefont{{Charbonnel}}},
  \bibinfo{author}{\bibfnamefont{S.}~\bibnamefont{{Talon}}}, \bibnamefont{and}
  \bibinfo{author}{\bibfnamefont{S.}~\bibnamefont{{Ekstr{\"o}m}}},
  \bibinfo{journal}{\apss} \textbf{\bibinfo{volume}{316}}, \bibinfo{pages}{43}
  (\bibinfo{year}{2008}).

\bibitem[{\citenamefont{{Ekstr{\"o}m} et~al.}(2012)\citenamefont{{Ekstr{\"o}m},
  {Georgy}, {Eggenberger}, {Meynet}, and et~al.}}]{Ekstrom2012}
\bibinfo{author}{\bibfnamefont{S.}~\bibnamefont{{Ekstr{\"o}m}}},
  \bibinfo{author}{\bibfnamefont{C.}~\bibnamefont{{Georgy}}},
  \bibinfo{author}{\bibfnamefont{P.}~\bibnamefont{{Eggenberger}}},
  \bibinfo{author}{\bibfnamefont{G.}~\bibnamefont{{Meynet}}}, \bibnamefont{and}
  \bibinfo{author}{\bibnamefont{et~al.}}, \bibinfo{journal}{\aap}
  \textbf{\bibinfo{volume}{537}}, \bibinfo{eid}{A146} (\bibinfo{year}{2012}),
  \eprint{1110.5049}.

\bibitem[{\citenamefont{{Desai} et~al.}(2019)\citenamefont{{Desai}, {Metzger},
  and {Foucart}}}]{desai19}
\bibinfo{author}{\bibfnamefont{D.}~\bibnamefont{{Desai}}},
  \bibinfo{author}{\bibfnamefont{B.~D.} \bibnamefont{{Metzger}}},
  \bibnamefont{and}
  \bibinfo{author}{\bibfnamefont{F.}~\bibnamefont{{Foucart}}},
  \bibinfo{journal}{\mnras} \textbf{\bibinfo{volume}{485}},
  \bibinfo{pages}{4404} (\bibinfo{year}{2019}), \eprint{1812.04641}.

\bibitem[{\citenamefont{{Radice} et~al.}(2018)\citenamefont{{Radice}, {Perego},
  {Zappa}, and {Bernuzzi}}}]{radice18}
\bibinfo{author}{\bibfnamefont{D.}~\bibnamefont{{Radice}}},
  \bibinfo{author}{\bibfnamefont{A.}~\bibnamefont{{Perego}}},
  \bibinfo{author}{\bibfnamefont{F.}~\bibnamefont{{Zappa}}}, \bibnamefont{and}
  \bibinfo{author}{\bibfnamefont{S.}~\bibnamefont{{Bernuzzi}}},
  \bibinfo{journal}{\apj} \textbf{\bibinfo{volume}{852}}, \bibinfo{eid}{L29}
  (\bibinfo{year}{2018}), \eprint{1711.03647}.

\bibitem[{\citenamefont{{Riley} et~al.}(2019)\citenamefont{{Riley}, {Watts},
  {Bogdanov}, {Ray}, and et~al.}}]{Riley2019}
\bibinfo{author}{\bibfnamefont{T.~E.} \bibnamefont{{Riley}}},
  \bibinfo{author}{\bibfnamefont{A.~L.} \bibnamefont{{Watts}}},
  \bibinfo{author}{\bibfnamefont{S.}~\bibnamefont{{Bogdanov}}},
  \bibinfo{author}{\bibfnamefont{P.~S.} \bibnamefont{{Ray}}}, \bibnamefont{and}
  \bibinfo{author}{\bibnamefont{et~al.}}, \bibinfo{journal}{\apjl}
  \textbf{\bibinfo{volume}{887}}, \bibinfo{eid}{L21} (\bibinfo{year}{2019}),
  \eprint{1912.05702}.

\bibitem[{\citenamefont{{Sarin} et~al.}(2021)\citenamefont{{Sarin}, {Ashton},
  {Lasky}, {Ackley}, {Mong}, and {Galloway}}}]{Sarin2021_cdf}
\bibinfo{author}{\bibfnamefont{N.}~\bibnamefont{{Sarin}}},
  \bibinfo{author}{\bibfnamefont{G.}~\bibnamefont{{Ashton}}},
  \bibinfo{author}{\bibfnamefont{P.~D.} \bibnamefont{{Lasky}}},
  \bibinfo{author}{\bibfnamefont{K.}~\bibnamefont{{Ackley}}},
  \bibinfo{author}{\bibfnamefont{Y.-L.} \bibnamefont{{Mong}}},
  \bibnamefont{and} \bibinfo{author}{\bibfnamefont{D.~K.}
  \bibnamefont{{Galloway}}}, \bibinfo{journal}{arXiv e-prints}
  \bibinfo{eid}{arXiv:2105.10108} (\bibinfo{year}{2021}), \eprint{2105.10108}.

\bibitem[{\citenamefont{{Hayes} et~al.}(2020)\citenamefont{{Hayes}, {Heng},
  {Veitch}, and {Williams}}}]{Hayes2020}
\bibinfo{author}{\bibfnamefont{F.}~\bibnamefont{{Hayes}}},
  \bibinfo{author}{\bibfnamefont{I.~S.} \bibnamefont{{Heng}}},
  \bibinfo{author}{\bibfnamefont{J.}~\bibnamefont{{Veitch}}}, \bibnamefont{and}
  \bibinfo{author}{\bibfnamefont{D.}~\bibnamefont{{Williams}}},
  \bibinfo{journal}{\apj} \textbf{\bibinfo{volume}{891}}, \bibinfo{eid}{124}
  (\bibinfo{year}{2020}), \eprint{1911.04190}.

\bibitem[{\citenamefont{{Kumar} and {Zhang}}(2015)}]{kumar15}
\bibinfo{author}{\bibfnamefont{P.}~\bibnamefont{{Kumar}}} \bibnamefont{and}
  \bibinfo{author}{\bibfnamefont{B.}~\bibnamefont{{Zhang}}},
  \bibinfo{journal}{\physrep} \textbf{\bibinfo{volume}{561}},
  \bibinfo{pages}{1} (\bibinfo{year}{2015}), \eprint{1410.0679}.

\bibitem[{\citenamefont{{Yang} et~al.}(2017)\citenamefont{{Yang}, {Valenti},
  {Cappellaro}, {Sand}, {Tartaglia}, {Corsi}, {Reichart}, {Haislip}, and
  {Kouprianov}}}]{Yang2017}
\bibinfo{author}{\bibfnamefont{S.}~\bibnamefont{{Yang}}},
  \bibinfo{author}{\bibfnamefont{S.}~\bibnamefont{{Valenti}}},
  \bibinfo{author}{\bibfnamefont{E.}~\bibnamefont{{Cappellaro}}},
  \bibinfo{author}{\bibfnamefont{D.~J.} \bibnamefont{{Sand}}},
  \bibinfo{author}{\bibfnamefont{L.}~\bibnamefont{{Tartaglia}}},
  \bibinfo{author}{\bibfnamefont{A.}~\bibnamefont{{Corsi}}},
  \bibinfo{author}{\bibfnamefont{D.~E.} \bibnamefont{{Reichart}}},
  \bibinfo{author}{\bibfnamefont{J.}~\bibnamefont{{Haislip}}},
  \bibnamefont{and}
  \bibinfo{author}{\bibfnamefont{V.}~\bibnamefont{{Kouprianov}}},
  \bibinfo{journal}{\apjl} \textbf{\bibinfo{volume}{851}}, \bibinfo{eid}{L48}
  (\bibinfo{year}{2017}), \eprint{1710.05864}.

\bibitem[{\citenamefont{{Andreoni} et~al.}(2021)\citenamefont{{Andreoni},
  {Coughlin}, {Kool}, {Kasliwal}, and et~al.}}]{Andreoni2021}
\bibinfo{author}{\bibfnamefont{I.}~\bibnamefont{{Andreoni}}},
  \bibinfo{author}{\bibfnamefont{M.~W.} \bibnamefont{{Coughlin}}},
  \bibinfo{author}{\bibfnamefont{E.~C.} \bibnamefont{{Kool}}},
  \bibinfo{author}{\bibfnamefont{M.~M.} \bibnamefont{{Kasliwal}}},
  \bibnamefont{and} \bibinfo{author}{\bibnamefont{et~al.}},
  \bibinfo{journal}{\apj} \textbf{\bibinfo{volume}{918}}, \bibinfo{eid}{63}
  (\bibinfo{year}{2021}), \eprint{2104.06352}.

\end{thebibliography}
\end{document}